\documentclass[format=acmsmall, review=false, screen=true]{acmart}

\usepackage{booktabs} 

\usepackage[ruled]{algorithm2e} 

\SetAlFnt{\small}
\SetAlCapFnt{\small}
\SetAlCapNameFnt{\small}
\SetAlCapHSkip{0pt}
\IncMargin{-\parindent}

\usepackage{multirow} 
\usepackage{rotating} 
\usepackage{paralist}
\usepackage{todonotes}
\usepackage{ulem}
\newcommand*\rot{\rotatebox{90}}

\setcopyright{rightsretained}

\begin{document}
\title[Status Quo in Requirements Engineering]{Status Quo in Requirements Engineering: \\ A Theory and a Global Family of Surveys}

\author{Stefan Wagner}
\orcid{0000-0002-5256-8429}
\affiliation{%
  \institution{University of Stuttgart}
  \city{Stuttgart}
  \country{Germany}}
\email{stefan.wagner@iste.uni-stuttgart.de}
\author{Daniel M\'{e}ndez Fern\'{a}ndez}
\affiliation{%
  \institution{Technical University of Munich}
  \city{Garching}
  \country{Germany}
}
\email{daniel.mendez@tum.de}
\author{Michael Felderer}
\affiliation{%
 \institution{University of Innsbruck}
 \city{Innsbruck}
 \country{Austria}}
\affiliation{%
 \institution{Blekinge Institute of Technology}
 \city{Karlskrona}
 \country{Sweden}}
\email{michael.felderer@uibk.ac.at}
\author{Antonio Vetr\`o}
\affiliation{%
  \institution{Nexa Center for Internet \& Society, DAUIN, Politecnico di Torino}
  \city{Torino}
  \country{Italy}
}
\email{antonio.vetro@polito.it}
\author{Marcos Kalinowski}
\affiliation{%
  \institution{Pontifical Catholic University of Rio de Janeiro}
  \city{Rio de Janeiro}
  \country{Brazil}}
\email{kalinowski@inf.puc-rio.br}
\author{Roel Wieringa}
\affiliation{%
  \institution{University of Twente}
  \city{Enschede}
  \country{The Netherlands}
}
\email{r.j.wieringa@utwente.nl}
\author{Dietmar Pfahl}
\affiliation{%
  \institution{University of Tartu}
  \city{Tartu}
  \country{Estonia}
}
\email{dietmar.pfahl@ut.ee}
\author{Tayana Conte}
\affiliation{%
  \institution{Universidade Federal do Amazonas}
  \city{Manaus}
  \country{Brazil}
}
\email{tayanaconte@gmail.com}
\author{Marie-Therese Christiansson}
\affiliation{%
  \institution{Karlstad University}
  \city{Karlstad}
  \country{Sweden}
}
\email{marie-therese.christiansson@kau.se}
\author{Desmond Greer}
\affiliation{%
  \institution{Queen's University Belfast}
  \city{Belfast}
  \country{UK}
}
\email{des.greer@qub.ac.uk}

\author{Casper Lassenius}
\affiliation{%
  \institution{Aalto University}
  \city{Espoo}
  \country{Finland}}
  \email{casper.lassenius@aa.fi}
  \affiliation{%
  \institution{SimulaMet}
  \city{Oslo}
  \country{Norway}}
\email{casper@simula.no}

\author{Tomi M\"annist\"o}
\affiliation{%
  \institution{University of Helsinki}
  \city{Helsinki}
  \country{Finland} 
}
\email{tomi.mannisto@helsinki.fi}
\author{Maleknaz Nayebi}
\affiliation{%
  \institution{University of Calgary}
  \city{Calgary}
  \country{Canada}
}
\email{mnayebi@ucalgary.ca}
\author{Markku Oivo}
\affiliation{%
  \institution{University of Oulu}
  \city{Oulu}
  \country{Finland}
}
\email{markku.oivo@oulu.fi}
\author{Birgit Penzenstadler}
\affiliation{%
  \institution{California State University, Long Beach}
  \city{Long Beach}
  \country{USA}
}
\email{birgit.penzenstadler@csulb.edu}
\author{Rafael Prikladnicki}
\affiliation{%
  \institution{Pontif\'icia Universidade Cat\'olica do Rio Grande do Sul}
  \city{Porto Alegre}
  \country{Brazil}
}
\email{rafael.prikladnicki@gmail.com}
\author{Guenther Ruhe}
\affiliation{%
  \institution{University of Calgary}
  \city{Calgary}
  \country{Canada}
}
\email{ruhe@ucalgary.ca}
\author{Andr\'e Schekelmann}
\affiliation{%
  \institution{Hochschule Niederrhein}
  \city{Krefeld}
  \country{Germany}
}
\email{andre.schekelmann@hs-niederrhein.de}
\author{Sagar Sen}
\affiliation{%
  \institution{Simula}
  \city{Fornebu}
  \country{Norway}
}
\email{sagar@simula.no}
\author{Rodrigo Sp\'inola}
\affiliation{%
  \institution{Salvador University - UNIFACS}
  \city{Salvador}
  \country{Brazil}
}
\email{rodrigoospinola@gmail.com}
\author{Ahmed Tuzcu}
\affiliation{%
  \institution{zeb.rolfes.schierenbeck.associates GmbH}
  \city{Munich}
  \country{Germany}
}
\email{atuzcu@zeb.de}
\author{Jose Luis de la Vara}
\affiliation{%
  \institution{Carlos III University of Madrid}
  \city{Madrid}
  \country{Spain}
}
\email{jvara@inf.uc3m.es}

\author{Dietmar Winkler}
\affiliation{%
  \institution{Technische Universit\"at Wien, CDL-SQI}
  \city{Vienna}
  \country{Austria}
}
\email{dietmar.winkler@tuwien.ac.at}

\begin{abstract}
\textbf{Context:}
Requirements Engineering (RE) has established itself as a software engineering discipline over the past decades. While researchers have been investigating the RE discipline with a plethora of empirical studies, attempts to systematically derive an empirical theory in context of the RE discipline have just recently been started.
However, such a theory is needed if we are to define and motivate guidance in performing high quality RE research and practice.

\textbf{Objective:} We aim at providing an empirical and externally valid foundation for a theory of RE practice, which helps software engineers establish effective and efficient RE processes in a problem-driven manner.

\textbf{Method:} We designed a survey instrument and an engineer-focused theory that was first piloted in Germany and, after making substantial modifications, has now been replicated in 10 countries world-wide. We have a theory in the form of a set of propositions inferred from our experiences and available studies, as well as the results from our pilot study in Germany. We evaluate the propositions with bootstrapped confidence intervals and derive potential explanations for the propositions.

\textbf{Results:} In this article, we report on the design of the family of surveys, its underlying theory, and the full results obtained from the replication studies conducted in 10 countries with participants from 228 organisations. Our results represent a substantial step forward towards developing an empirical theory of RE practice. The results reveal, for example, that there are no strong differences between organisations in different countries and regions, that interviews, facilitated meetings and prototyping are the most used elicitation techniques, that requirements are often documented textually, that traces between requirements and code or design documents are common, that requirements specifications themselves are rarely changed and that requirements engineering (process) improvement endeavours are mostly internally driven.

\textbf{Conclusion:}
Our study establishes a theory that can be used as starting point for many further studies for more detailed investigations. Practitioners can use the results as theory-supported guidance on selecting suitable RE methods and techniques.
\end{abstract}

%
%

\ccsdesc[500]{General and reference~Empirical studies}
\ccsdesc[500]{Software and its engineering~Requirements analysis}

%
%

\keywords{Requirements Engineering, Theory, Survey Research, Replication}

\maketitle

\renewcommand{\shortauthors}{S.~Wagner et al.}

\section{Introduction}
\label{sec:Introduction}

As stated by Wohlin \textit{et al.}~\cite{Wohlin:2015:GTS:2827361.2827466}: ``There exists no generally accepted theory in software engineering, and at the same time a scientific discipline needs theories. Some laws, hypotheses and conjectures exist, but yet no generally accepted theory.''~\cite{Wohlin:2015:GTS:2827361.2827466} What is true for the whole discipline holds especially for its sub-disciplines, in particular for requirements engineering (RE). In a literature survey published in 2007, Hannay \textit{et al.}~\cite{HDT07} identified 103 publications  of the period 1993--2002 that report on software engineering experiments. Out of those only 24 publications used in total 40 theories to justify their research questions, explain the results or, rarely, test and modify theories. The authors also noticed that most theory has been developed in other disciplines and only a small number of theories detected were genuine to the software engineering discipline. This observation was also made by Endres and Rombach~\cite{endres2003} in their collection of empirical observations, laws and theories on software and systems engineering published in 2003. Moreover, most of the theories seem to relate to management tasks rather than requirements specification, technical design, verification and validation tasks.

The lack of a general theory of software engineering triggered the SEMAT  initiative \textit{(Software Engineering Method and Theory)} which, among other things, develops and maintains the ESSENCE standard aiming at providing comprehensive definitions and descriptions of the kernel and the language for software engineering methods. However, until today, all attempts to collect theories of software engineering or to develop a theory of software engineering have very little to offer with regards to RE, and if there is some theory developed, it often relates to RE management tasks (e.g.,~\cite{Ali201649,Knauss201685}). Therefore, Daniel M\'endez Fern\'andez
and Stefan Wagner started the NaPiRE initiative\footnote{\url{http://www.napire.org}} \textit{(Naming the Pain in Requirements Engineering)} in 2012 which constitutes a globally distributed family of surveys on RE. In accordance with the statement made by Stol and Fitzgerald~\citeN{stol2015theory} that theory is ``both a driver for, and a result of empirical research'', the goal of this initiative is to create an empirical theory on requirements engineering practices and problems. For this, we adopt the view that a \emph{theory} is a set of constructs and general propositions and, possibly, explanations of those propositions \cite{sjoberg2008building,Wier14}.

How RE is conducted in a given context is a crucial factor in successful development projects as the elicitation, specification, and validation of requirements
are critical determinants of software and system quality~\cite{broy06_mbRE}. At the same time, requirements engineering is highly volatile and inherently complex due to the involvement of interdisciplinary stakeholders and uncertainty about many aspects that are not clear at the beginning of a project. The diversity of how requirements engineering is performed in various industrial environments, each having its particularities in the domains of application or the software process models used, makes it difficult for industry to define and apply high-quality requirements engineering~\cite{MWLBC10}.

\subsection{Problem Statement}

From a researcher's perspective, the diversity of requirements engineering contexts makes it hard to develop general theories.
Rather, we expect to develop a few general propositions and a lot more context-specific propositions.
Empirical research in RE thereby becomes a crucial and challenging task. Empirical studies of all kinds, ranging from classical action research through observational studies to broad exploratory surveys, are necessary to understand the practical needs and improvement goals in requirements engineering to guide problem-driven research and to empirically validate new research proposals~\cite{CDW12}. Since requirements engineering, like most sub-disciplines of software engineering, is highly human-based, we face the challenge to create a solid empirical basis that allows for generalisations taking into account the human factors that influence the anyway hard to standardise discipline. To address this challenge, survey research has become an indispensable means to investigate RE across many contexts.

\subsection{Research Objective}

Our long-term research objective is to establish an empirically-based descriptive and explanatory theory on the status quo of requirements engineering practice allowing us to guide future research in a theory-based manner. According to \citeauthor{sjoberg2008building} \cite{sjoberg2008building}, ``Theory development consists of inductive and deductive aspects, and may be initiated from both the practical or from the theoretical realm''. In preparation of our first NaPiRE survey conducted in Germany in 2012/13, we deductively developed an initial, engineer-focused theory of requirements engineering practice based on experience gained during research collaborations with industry and ideas taken from the scientific literature~\cite{MW14}. 

For this article, we improved the theory based on the results of the first run of the survey and extended it by further propositions on requirements elicitation, documentation and test alignment to better cover the whole requirement engineering process. We reflected these changes in a new version of the survey instrument and partially replicated the first survey in 10 countries. We report on this partial replication in this article and call it ``second run'' in the following. Our goal here is to use the results of this replication to evaluate and further improve the RE theory. In ~\cite{napire:emse15} important problems, causes, and effects have been reported. Among others, \textit{poor requirements elicitation techniques} and  \textit{missing completeness checks} have been identified as important causes that lead to requirements engineering problems. In this article we focus on RE standards and their application, requirements elicitation approaches, and RE improvement options. Specifically, we want to answer the following research questions: (i) how are requirements elicited and documented? (ii) how are requirements changed and aligned with tests? (iii) why and how are RE standards applied and tailored? and (iv) how is RE improved?

\subsection{Contribution}
\label{sec:contribution}

In~\cite{MW14}, we published the initial theory as well as the design of the used survey instrument and discussed preliminary results from the first run conducted in Germany. We presented and analysed the qualitative data regarding problems in practice from the second run of the survey without relating it to the prior theory in \cite{napire:emse15}. In the present article, we evaluate and propose improvements to the theory of RE practice based mostly on the quantitative analysis of the answers on the
status quo from the second run. More in particular, we present the following contributions:
\begin{compactenum}
\item
We substantially enhanced our initial theory after the analysis of the results gained from the first run in Germany and using input from collaborating researchers received during a thematic workshop held within the ISERN \textit{(International Software Engineering Network)} meeting in 2015.
The resulting theory includes for each research question a set of propositions concerning requirements elicitation, documentation, change and alignment, standards and improvement with a focus on the involved engineers. We use the propositions in our theory to test our results during our analysis procedure.
\item We report the full results from 10 countries including the responses from 228 organisations and a detailed analysis of those results via a calculation of confidence intervals with respect to our theory. This allows a statistically sound interpretation of the answers and a validation of the theory.
\item Based on the quantitative results, additional qualitative answers by the respondents and further interpretation, we propose corresponding improvements of the underlying theory in the form of changed or new propositions and explanations.
\end{compactenum}

Our contribution is thus intended to serve both RE practitioners interested in the status quo of RE practice and RE researchers interested in theories that aim to describe real-world phenomena of RE practice.

\subsection{Outline}

We structure the remainder of this article as follows: We start with a discussion of the related work in Section~\ref{sec:RelatedWork} on theories and survey research in requirements engineering as well as the background on the NaPiRE initiative and then present our current theory on the status quo in requirements engineering practice in Section~\ref{sec:Theory}. In Section~\ref{sec:Design}, we present the design of the survey which we use to validate the theory. We discuss the results of the survey in Section~\ref{sec:Results} structured by the research questions. Finally, we summarise and discuss implications and future work in Section~\ref{sec:Conclusion}.

\section{Related Work}
\label{sec:RelatedWork}

In this section we discuss theories on requirements engineering, survey research related to the scope of our contribution, and introduce the NaPiRE initiative as well as previously published material in this context in more detail.\footnote{Parts of section~\ref{sec:survesresearchonre} and section~\ref{sec:NaPiREInitiative} are based on our related work discussion in \cite{MW14} as the related work has not changed significantly. It has also been used in \cite{napire:emse15}.}

\subsection{Theories on Requirements Engineering}
\label{sec:TheoriesinRE}

Sj{\o}berg \textit{et al.}~\cite{sjoberg2008building} proposed a number of activities to define software engineering theories, namely (1) defining the constructs of the theory, (2) defining the propositions of the theory, (3) providing explanations to justify the theory, (4) determining the scope of the theory, as well as (5) testing the theory through empirical research. In our work, we follow these proposed steps.

The literature in the broader area of software engineering provides a broad spectrum on theories. In 2009, for instance, Jacobson~\cite{jacobson2009we} motivated the need for a theory in software engineering initiating the SEMAT initiative \textit{(Software Engineering Method and Theory)}~\cite{jacobson2009semat}. According to Jacobson, software engineering is gravely hampered by (1) the prevalence of fads more typical of fashion industry than of an engineering discipline; (2) the lack of a sound, widely accepted theoretical basis; (3) the huge number of methods and method variants with differences little understood and artificially magnified; (4) the lack of credible experimental evaluation and validation; and finally (5) the split between industry practice and academic research. NaPiRE aims to improve on all these aspects in the area of requirements engineering. Tarpit~\cite{johnson2016tarpit} is an example of a more general theory of software engineering derived from four underlying theoretical fields, i.e., (1) languages and automata, (2) cognitive architecture, (3) problem solving, and (4) organisation structure. Its applicability was explored in the cases of Brooks' law, domain specific languages and continuous integration. Yet, it has not been applied to RE so far. 

Bjarnason \textit{et al.}~\cite{bjarnason2016theory} propose a theory of distances in software engineering. The theory is based on a mapping study of RE distances~\cite{bjarnason2013distances} as well as RE and testing challenges and practices~\cite{bjarnason2014challenges}. The theory of distances in software engineering considers people-related distances but also artefact-related distances, whereas our theory focuses on artefact-related distances but also considers other aspects like RE documentation, standards and improvement.

Theories on requirements engineering are even more rare. The SWEBOK~\cite{bourque2014guide} aims to describe the body of knowledge in software engineering including a knowledge area on software requirements. Yet, the SWEBOK does not constitute an empirically validated theory on RE, but it is created by consensus of the participating people. While we consider it an important effort and starting point, our theory contains more details on the RE process and methods including explicit propositions, which we test empirically, as well as explanations.

The \textit{handbook of software and systems engineering}~\cite{endres2003} is a rare exception reporting on empirical observations, laws, and theories in various fields including requirements engineering. Theories cover, for example, that requirements deficiencies --  in particular, incomplete, incorrect and volatile requirements -- are the prime sources of project failures. They became part of the problem analysis in \cite{napire:emse15} but are so far not integrated in the theory in the current article.

Besides those well known, mostly descriptive and causal, theories, there is still a lack of theories in requirements engineering. This article provides a step towards closing this gap. In particular, we aim for a descriptive and explanatory theory validated by a broad survey. Therefore, we discuss work related to survey research on requirements engineering next.

\subsection{Survey Research on Requirements Engineering}
\label{sec:survesresearchonre}
In RE survey research, there have been investigations of techniques and methods and investigations of general practices. We will discuss the most relevant ones in the following.

Contributions that investigate techniques and methods analyse, for example, selected requirements phases and which techniques are suitable to support typical tasks in those phases.
Cox, Niazi and Verner~\cite{CNV09} performed a broader investigation of all phases to analyse the perceived value of the RE practices recommended by \citeN{Sommerville1997}.
Studies like those reveal the effects of given techniques when applying them in practical contexts.

A similar focus, but exclusively narrowed down to the area of RE, had the study of Kamata and Tamai~\cite{IT07}. They analysed the criticality of the single parts of the IEEE software requirements specification Std.~830-1998~\cite{IEEE830} for project success. Palomares, Quer and Franch~\cite{palomares2017requirements} investigated the use of requirements reuse and requirements patterns with a survey among practitioners. They found that reuse and patterns are done only by a minority regularly. We do not cover requirements reuse in our survey.

Nikula, Sajaniemi and K\"alvi\"ainen~\cite{nikula2000sps} present a survey on RE at the organisational level of small and medium-sized companies in Finland. Based on their findings, they inferred improvement goals, e.g., for optimising knowledge transfer.
Staples \textit{et al.}~\cite{SNJABR07} conducted a study investigating the industrial reluctance to software process improvement.
They discovered different reasons why organisations do not adopt normative improvement solutions, for example, CMMI
and related frameworks (focussing on assessing and benchmarking companies rather than on problem-driven improvements
\cite{NMJ09,PIGO08}). 
Example reasons for a reluctance to normative improvement frameworks were the small company size because of which the respondents did not see clear benefit. 

Neil and Laplante~\citeN{Neill:2003ui} conducted in 2003 a survey in the USA with a focus on some aspects we also cover here.
In particular, they found that most of the respondents use scenarios and use cases for requirements elicitation followed by focus groups
and informal modelling. Separately, they also found that 60\% of the respondents do prototyping.
Kassab, Neill and Laplante~\citeN{Kassab2014} updated this survey in a similar manner. Brainstorming, interviews and user stories were techniques most mentioned by their respondents, while workshops were only mentioned by roughly 20~\%.

Although giving valuable insights into industrial environments, the discussed surveys do not give a comprehensive picture
as they focus on single aspects in RE, such as problems in RE processes or RE improvements, or they focus on specific countries. 
To close this gap in literature, we designed a family of surveys. The design of the survey as well as the interpretation of the results are both conducted along an initial theory~\cite{MW14}. By relying on an initial theory built on the basis of our experiences and available studies, and by bringing together different interdisciplinary communities during replications, the family of surveys shall contribute to an empirical basis for theories and problem-driven research in RE. The replications around the world give us a more heterogeneous sample. If some phenomena stand out nevertheless, then they are phenomena occurring in heterogeneous contexts, which have at least different cultures and languages.

Furthermore, there are also several literature studies available that survey the scientific literature on specific requirements engineering activities and are related to the activities considered in this article (i.e., requirements elicitation, requirements documentation, requirements change management, requirements test alignment, requirements engineering process standard and requirements engineering improvement). Dieste and Juristo~\cite{dieste2011systematic} provide a systematic review on empirical studies on requirements elicitation techniques, Condori-Fernandez et al.~\cite{condori2009systematic} a systematic mapping on empirical evaluation of software requirements specification techniques, Inayat et al.~\cite{inayat2015systematic} a systematic review on agile requirements engineering practices and challenges, Barmi et al.~\cite{barmi2011alignment} a systematic mapping on requirements specification and testing alignment, Loniewski et al.~\cite{loniewski2010systematic} a systematic review on the use of requirements engineering techniques in model-driven development and finally Pekar et al.~\cite{pekar2014improvement} a systematic mapping on improvement methods for requirements specification.
They provide an interesting counterpart to our view into practice.

\subsection{The NaPiRE Initiative}
\label{sec:NaPiREInitiative}  

The basic idea of NaPiRE was to establish a broad survey investigating the status quo of requirements engineering in practice
together with contemporary problems practitioners encounter. This should provide an empirical basis for theories about RE and should lead to the identification of interesting further research areas.

When we started NaPiRE, we were convinced that because of the diversity of RE in research and practice, we would not be able to achieve this ambitious goal in a small team and in a single survey. Therefore, NaPiRE was created as a means to collaborate with researchers from all over the world to conduct the survey in different countries and in a replicated manner. This allows us to investigate RE in different cultural environments and increase the overall sample size while covering slightly different areas over time and while also having the possibility to observe trends. 

More precisely, we started NaPiRE by establishing a first set of theory patterns based on isolated work published in the research community. With each replication, our hope is to further strengthen this initial theory by corroborating it based on the gathered data and modifying it where we have no explanation for the observed phenomena (e.g. by removing initial hypotheses or by changing them). That is to say, based on the results from each replication, we slightly adapt the questions in the questionnaire as well as the answer options for the next replication. 
The conduct of NaPiRE is generally guided by the four principles described in Tab.~\ref{tab:principles}.

\begin{table}[htb]
\caption{Guiding principles of NaPiRE\label{tab:principles}}{%
\begin{tabular}{r p{0.65\linewidth}}
\toprule
\textbf{Openness} & Openness begins by cordially inviting researchers and practitioners of any software-engineering-related community
to contribute to NaPiRE and ends by disclosing our results and reports without any restrictions or commercial interest. \\
\textbf{Transparency} & All results obtained from the distributed surveys are committed to an open repository. This shall allow
other researchers an independent data analysis and interpretation.\\
\textbf{Anonymity} & The participation in NaPiRE in the form of a survey respondent is possible by invitation only. This
supports a transparent result set and response rate. We collect no personal data, however, and every
questionnaire obtained from the survey will be carefully cleansed of information that might be traced back to a specific
company to ensure that no personal data will be disclosed to the public. That is, we guarantee that no answer set
can be related to survey participants and their organisations.\\
\textbf{Accuracy and Validity} & With accuracy and validity, we refer in particular to the data collection and to the data
analysis. Each question in the survey is carefully defined according to a jointly defined theory to specifically confirm
or refute existing expectations. The data analysis is furthermore performed in joint collaboration by different researchers
to maximise the validity of the results.\\
\bottomrule
\end{tabular}}
\end{table}

To organize our work, e.g., on the NaPiRE instrumentation, we hold joint community workshops. These community workshops take so far place at the annual meetings of the International Software Engineering Research Network, which forms part of the annual Empirical Software Engineering International Week. We use these workshops biannually to organize the modification of the questionnaires as a preparation for one NaPiRE run and the next workshop to discuss the results obtained from the run before entering the next round. In the workshops where we prepare one replication, we discuss to which extent our initial expectations (again, based on literature) are reflected by the actual results from the previous run and, in cases where the results deviate from our expectations, whether other propositions should be included instead. We jointly adapt the questionnaire (questions and answer options) before centrally implementing the questionnaire, organizing pilot studies with partners from the industry, and starting the distributed data collection.

\begin{figure}[!htb]
\centering
\includegraphics[width=.6\textwidth]{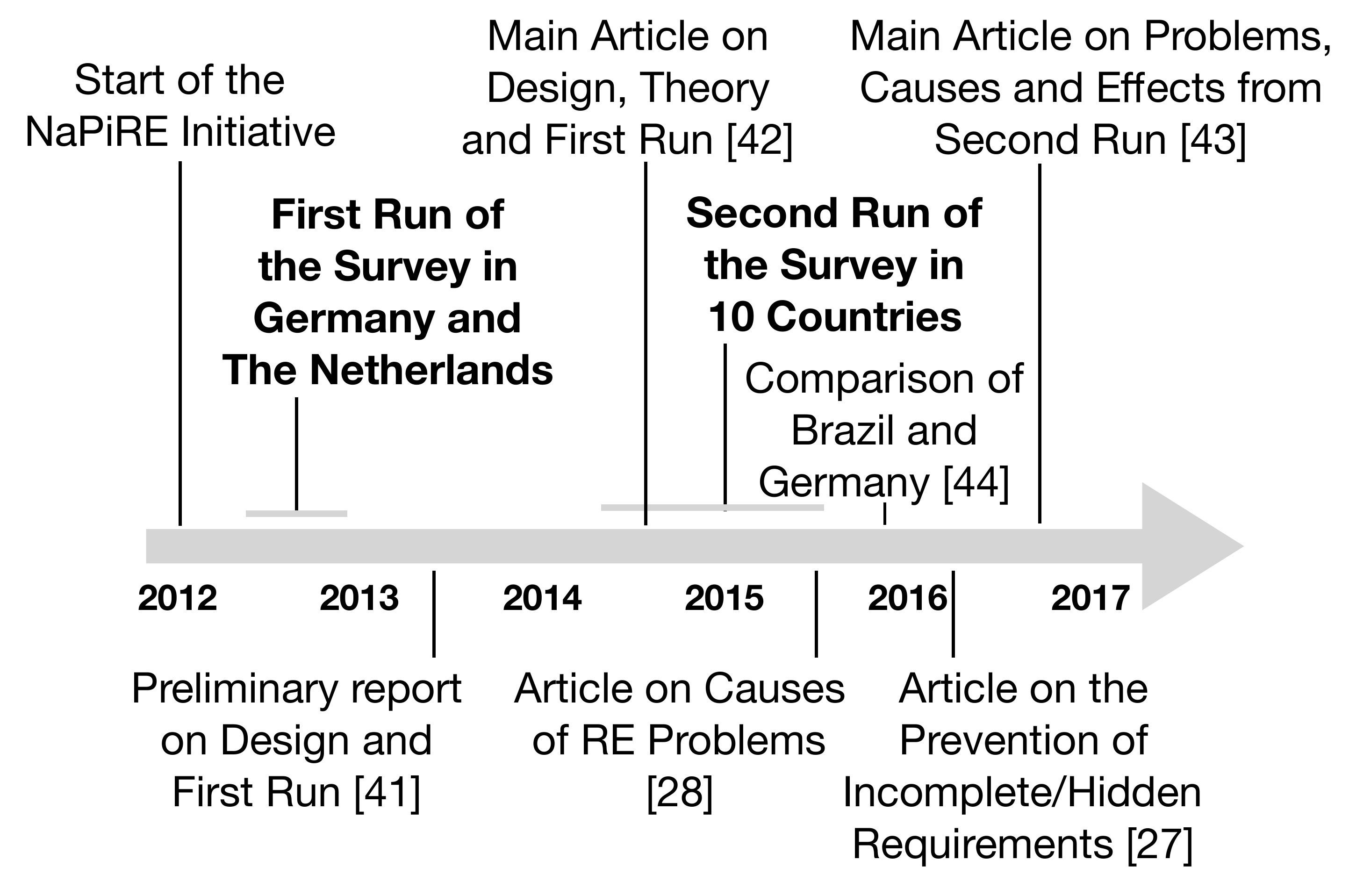}
\caption{Timeline of the major NaPiRE activities}\label{fig:napire-overview}
\end{figure}

At present, the NaPiRE initiative has members from 25 countries mostly from Europe, North-America, South-America and Asia. There have been two completed runs of the survey and the third is taking place at the time of writing this manuscript. An overview of the timeline of the activities of NaPiRE so far is shown in Figure~\ref{fig:napire-overview}. The first was the test run performed only in Germany and the Netherlands in 2012/13. The second run was performed in 10 countries in 2014/15. This article reports on the results from this second run. All up-to-date information on NaPiRE together with links to all publications, the instruments used for each replication, links to the published open data sets, as well as the steering manifesto of the initiative is available on the web site \url{http://www.napire.org}.

A preliminary version of the report on the first run was published in \cite{MW2013a} and the detailed 
data and descriptive analysis is available as a technical report~\cite{MW13b}. This first run already covered the spectrum of status quo and problems. We described the study design with the bi-yearly replications and world-wide distribution in detail in the main article for the first run \cite{MW14}. It includes a first version of a theory of the status quo and problems in RE in the form of hypotheses. Overall, we were able to get full responses from 58 organisations to test the theory. We could support most of the proposed theory and discussed changes based on the data. We also made a detailed qualitative analysis of the experienced problems and how they manifest themselves. The current article extends and improves the theory from the first run and evaluates it using mostly quantitative analysis.

For the publications of the second run, we so far have concentrated on specific aspects and/or the data from only one or two countries. None of these has described the full status quo nor did they present or evaluate a theory. In \cite{kalinowski:seke15}, we used the Brazilian data to explore how to analyse problems and causes in RE in detail. Thereafter, in \cite{mendez:softw15}, we concentrated on analysing the similarities and differences in the experienced problems between Brazil and Germany. Our key insights in that paper were that the dominating factors are related to human interactions independent of country, project type or company. Furthermore, we observed a higher inclination to standardised development process models in Brazil and slightly more non-agile, plan-driven RE in Germany.

In \cite{kalinowski:swqd16}, we focused on the often mentioned problem of \emph{Incomplete and/or hidden requirements} and investigated common causes for this problem based on the Austrian and Brazilian data. The most common causes we found were \emph{Missing qualification of RE team members}, \emph{Lack of experience}, \emph{Missing domain knowledge}, \emph{Unclear business needs} and \emph{Poorly defined requirements}.

In \cite{wagner2018agile}, we report on the status quo and critical problems of agile requirements engineering. The study shows that the backlog is the central means to deal with changing requirements, traces between requirements and code are explicitly managed, and testing and RE are typically aligned. Furthermore, continuous improvement of RE is performed due to intrinsic motivation and RE standards are commonly practiced. Main problems of agile requirements engineering with critical consequences are incomplete requirements, communication flaws and moving targets.

Finally, we describe the analysis of the part of the questionnaire on contemporary problems, their causes and effects in the corresponding main article \cite{napire:emse15}. In the current article, we use the same data set (and therefore the same context factors such as application domains) as in \cite{kalinowski:seke15,kalinowski:swqd16,napire:emse15}. Yet, we use a part of the data set not covered in the previous papers: the parts on the participants use of practices in requirements elicitation, requirements documentation, requirements change and alignment and their use of requirements engineering standards as well as their improvement.

\section{A Theory on the Status Quo in Requirements Engineering}
\label{sec:Theory}

The basis of our family of studies is a theory on the status quo of requirements engineering practice. We started the
theory in the first round of the studies and documented it in \cite{MW14}.
By \emph{theory} we mean a set of constructs and general propositions and, possibly, explanations of those propositions \cite{sjoberg2008building,Wier14}.
Each theory has a scope, and our scope is the world-wide practice of requirements engineering.
The theory is currently populated with general descriptive propositions about how requirements engineering is practiced in industry \cite{MW14} and about contemporary problems in requirements engineering \cite{napire:emse15}.
In this paper, we will add new propositions as well as explanations to the theory.

We removed the part of the initial theory about the expectations of practitioners on good requirements engineering practices, as it did not give us particularly interesting insights. We moved, however, the propositions we found useful to the part on requirements engineering standards. Furthermore, we restructured the theory into the following parts:

\begin{itemize}
\item Requirements Elicitation
\item Requirements Documentation
\item Requirements Change and Alignment
\item Requirements Engineering Standards
\item Requirements Engineering Improvement
\end{itemize}

The initial theory contained hypotheses for all these parts but with differing emphasis. We especially detailed the theory in
the area of documentation to better capture what techniques are used for what. Moreover, we added hypotheses on aligning
requirements and tests which has not been investigated in our first study. The added hypotheses are based on our joint
understanding and observations from industry. We have not added any hypotheses based on the correlation analysis
from \cite{MW14} although that was its goal. The found correlations were all in the expectations on good requirements 
engineering practices, which we removed from the theory, or between different barriers for adopting RE process standards. All in
all, they were not particularly insightful.

For this second, restructured and extended theory, we decided to follow the proposal on how to document and structure
a software engineering theory by Sj{\o}berg \textit{et al.}~\cite{sjoberg2008building}. They propose to structure such theories
into the used \emph{constructs}, \emph{propositions} about relationships between the constructs and, finally, possible
\emph{explanations} for the propositions. We describe constructs and an overview of the propositions in this section. We then provide detailed propositions and explanations when we analyse the results of the survey in Section~\ref{sec:Results}. ``The difference between a proposition and an explanation is that the former is a relationship among constructs, and the latter is a relationship among constructs and other categories, which are not central enough to become constructs [\ldots].''~\cite{sjoberg2008building}

\begin{figure}[htbp]
\begin{center}
\includegraphics[width=\textwidth]{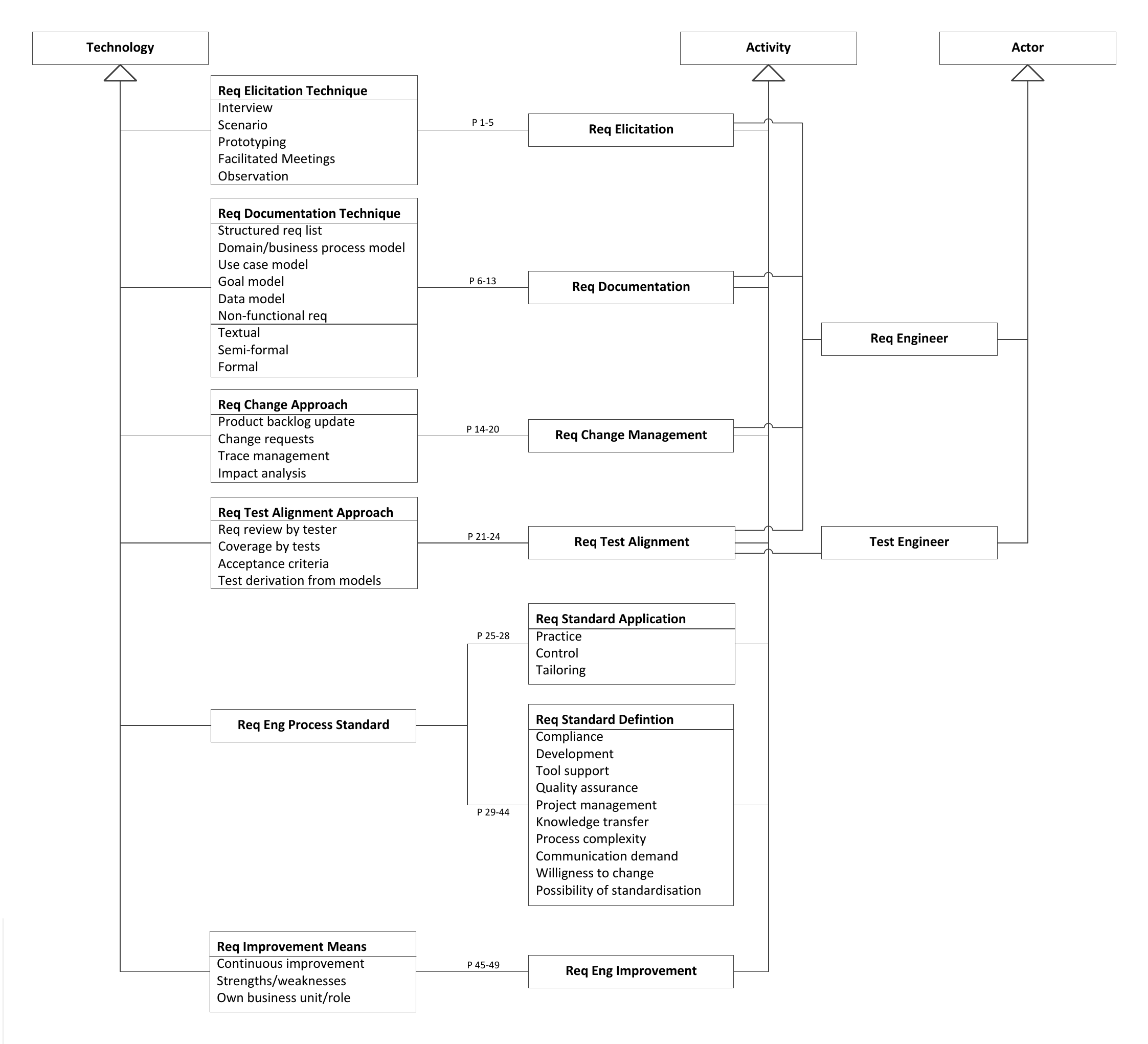}
\caption{Theory Overview: Constructs and index of propositions.}
\label{fig:theory-overview}
\end{center}
\end{figure}

Sj{\o}berg \textit{et al}.\ also propose a graphical representation inspired by UML class diagrams. We use that for an overview of the the main constructs and relationships in form of propositions in Figure~\ref{fig:theory-overview}. All main constructs of our theory are summarised in Table~\ref{tab:constructs} which also makes the scope of our theory explicit. We expect it to be applicable world-wide and cross-domain. The selection of activities, actors, and technologies is in line with the research questions with focus on requirements elicitation and documentation, requirements changes and testing, RE standards, and RE improvement options.

\begin{table}[htbp]
\caption{Main constructs and scope of the theory}{%
\begin{tabular}{l p{9cm} l}
\toprule
Constructs & & Type\\
\midrule
C~1 & Requirements Elicitation & Activity\\
C~2 & Requirements Documentation & Activity\\
C~3 & Requirements Change Management & Activity\\
C~4 & Requirements Test Alignment & Activity\\
C~5 & Requirements Standard Application & Activity\\
C~6 & Requirements Standard Definition & Activity\\
C~7 & Requirements Engineering Improvement & Activity\\
C~8 & Requirements Engineer & Actor\\
C~9 & Test Engineer & Actor\\
C~10 & Requirements Elicitation Technique & Technology\\
C~11 & Requirements Documentation Technique & Technology\\
C~12 & Requirements Change Approach & Technology\\
C~13 & Requirements Test Alignment Approach & Technology\\
C~14 & Requirements Engineering Process Standard & Technology\\
C~15 & Requirements Improvement Means & Technology\\
\midrule
Scope\\
\midrule
 & The theory is supposed to be applicable to contemporary requirements engineering
     in practice world-wide. There could be differences in different regions
     of the world because of cultural differences or different economic environments as
     well as differences in different application domains.\\
\bottomrule
\end{tabular}}
\label{tab:constructs}
\end{table}%

As suggested by Sj{\o}berg \textit{et al}., we structure the constructs into \emph{technology}, \emph{activity} and \emph{actors}. The main actors we describe in our theory are \emph{requirements engineers} and \emph{test engineers}. Most of the theory actually relates to the requirements engineer, but we found that the test engineer might play an important role as well. So far, we have not included explicit propositions involving the customers or users of the system. While this would be interesting given the important part they play in requirements engineering, the theory is already at the limit of what we could handle in a single survey. The theory should, however, be extended with these roles in the future.

We classified the propositions in terms of six activities related to requirements engineering: \emph{requirements elicitation}, \emph{requirements documentation}, \emph{requirements change management}, \emph{requirements test alignment}, \emph{requirements
engineering process standard} and \emph{requirements engineering improvement}. Elicitation and documentation form core activities in requirements engineering. Furthermore, requirements change over time, which gives rise to the activity of change management. 
To facilitate requirements-based testing, we included the activity of requirements-test alignment. All activities and related artefacts in requirements engineering are usually captured in a process standard containing a blueprint of the (idealised) way of working in a specific setting (a company, a team or a project). Finally, the process of requirements engineering should be regularly improved as any other part of software development processes.

We will describe the concrete propositions classified according to these activities together with the results and refined propositions and explanations in Section~\ref{sec:Results}. 
Please note that all propositions state that a technology or activity is used in practice. For all these propositions, we mean that the technologies or activities are \emph{commonly} in use. In particular, just because there are infrequent applications in practice, we would not state that \emph{it is used}. It has to cross a threshold that we will define in the analysis procedure (Sec.~\ref{sec:DataAnalyis}).

\section{Survey Design}
\label{sec:Design}

The overall methodology of the NaPiRE initiative and how the survey instrument has been developed and continuously adapted has been described earlier in section~\ref{sec:NaPiREInitiative}, hence, we will not repeat it. Instead, we define the research questions that drove the second run of the NaPiRE study and that are relevant for the research methodology in this run, including the study instrument, analysis procedures, and validity.

\subsection{Research Questions}

In agreement with the structure of the theory presented in Section~\ref{sec:Theory}, we have the four research questions listed
in Tab.~\ref{tab:rqs}. 
These questions are descriptive, and each of them has an explanatory sequel.
For example, if we have an answer on RQ~1, then  the subsequent question is: ``Why are requirements elicited and documented this way?''
Answering these explanatory questions is part of the analysis of results in section~\ref{sec:Results}.

\begin{table}[htb]
\caption{Descriptive research questions. Each question has an explanatory sequel.\label{tab:rqs}}{%
\begin{tabular}{p{0.13\linewidth}p{0.81\linewidth}}
\toprule
\textbf{RQ~1} & How are requirements elicited and documented? \\
\textbf{RQ~2} & How are requirements changed and aligned with tests? \\
\textbf{RQ~3} & How are RE standards applied and tailored? \\
\textbf{RQ~4} & How is RE improved? \\
\bottomrule
\end{tabular}}
\end{table}

We designed the survey in a way that we can investigate the updated theory and judge the support for it,
change existing propositions or add new ones for the next run of the survey, and add explanations where possible. 

\subsection{Survey Instrument}
\label{sec:Instrument}

The full instrument and codebook is openly available in the open data set~\cite{Mendez2018}. We have in total 34 questions in the survey, out of which 24 are 
relevant for the research
questions in this article, which are listed in Table~\ref{tab:instrument}. The further 10 questions are about general problems in requirements engineering. We analysed the answers to these questions in \cite{napire:emse15}.
In the fourth column, for each question, we denote whether it is an open question or a closed one and for closed questions whether the answers are mutually exclusive single choice answers (SC) or multiple choice ones (MC).
Most of the closed multiple choice questions include a free text option, e.g., ``other'' so that the respondents can express company-specific deviations from standards we ask for.
 
We use Likert items (an ordinal scale of 5) and defined as the maximum value ``agree'' and as the minimum value ``disagree'' and the middle (``neutral''). The latter allows the respondents to make a selection when they have, for example, no opinion on the given answer options.

The survey questions have been presented to respondents as listed in Table~\ref{tab:instrument}. At the end of
the survey, the respondents could enter their email address and freely add any other aspect that remained not addressed in the survey. We use this information as input for future modifications of the instrument.

\begin{table*}[htb]
\centering \scriptsize
\caption{Questions (simplified and condensed). Closed questions require mutually exclusive single choice answers (SC) or multiple choice answers (MC).\label{tab:instrument}}{%
\begin{tabular}{crp{0.65\linewidth}l}
\toprule
\textbf{RQ}& \textbf{No.}& \textbf{Question} & \textbf{Type} \\ \midrule
--           & Q~1        & What is the size of your company? & Closed(SC)\\
            & Q~2           & Please describe the main business area and application domain. & Open\\
             & Q~3          & Does your company participate in globally distributed projects? & Closed(SC)\\
             & Q~4          &  In which country are you personally located? & Open \\
            & Q~5           & To which project role are you most frequently assigned? & Closed(SC) \\
            & Q~6           &  How do you rate your experience in this role? & Closed(SC)\\
            & Q~7           &  Which organisational role does your company take most frequently in your projects? & Closed(SC)\\
            & Q~8           &  Which process model do you follow (or a variation of it)? & Closed(MC)\\ \midrule
RQ~1  & Q~9            &  How do you elicit requirements?	 &  Closed(MC)  \\
            & Q~10           &  How do you document functional requirements?	 & Closed(MC)  \\
            & Q~11           &  How do you document non-functional requirements?		       & Closed(SC)  \\ \midrule
RQ~2   &   Q~21           &  How do you perform change management in your requirements engineering? & Closed(MC)\\
            & Q~12           &  How do you deal with changing requirements after the initial release?	& Closed(SC)  \\
             & Q~13          &  Which traces do you explicitly manage?	& Closed(MC) \\
            & Q~14           &  How do you analyse the effect of changes to requirements? & Closed(MC) \\
            & Q~15           &  How do you align the software test with the requirements? & Closed(MC)\\ \midrule

RQ~3  & Q~16           &  What RE standard have you established at your company? & Closed(MC) \\
            & Q~17           &  Which reasons do you agree with as a motivation to define a company standard for RE in your company?& Likert \\
            & Q~18           &  Which reasons do you see as a barrier to define a company standard for RE in your company? & Likert\\
             & Q~19          &  Is the requirements engineering standard mandatory and practised? & Closed(SC)  \\
            & Q~20           &  How do you check the application of your requirements engineering standard? & Closed(MC)\\
            & Q~22           &  How is your RE standard applied (tailored) in your regular projects?	 & Closed(MC) \\ \midrule
RQ~4  & Q~23           &  Is your RE continuously improved? & Closed(SC) \\
            & Q~24            &  Why do you continuously improve your requirements engineering? & Closed(MC)\\
\bottomrule
\end{tabular}}
\end{table*}

\subsection{Data Collection}

The survey has been  conducted by invitation, for two main reasons: (1) to have better control over its distribution 
among specific companies and (2) to improve the response rate.
The responses were anonymous allowing our respondents to freely share their experiences made within their respective company.

For each company, we invited one respondent as a representative of the company. In case of large companies involving several autonomous business units working each in a different industrial sector and application domain, we selected a representative of each unit. Therefore, we call our unit of analysis ``organisation'' as an umbrella for individual companies or business units of larger companies. For the data collection, the country representatives defined an invitation list including contacts from different companies and initiated the data collection independently as an own survey project. The invitation list overlaps for the German invitees with the list from the first run. Yet, we have no way to track the overlap in actual respondents to the survey.

All surveys relied on the same survey tool\footnote{We implemented the survey as a Web application using the \emph{Enterprise Feedback Suite}.} hosted and administrated by the representatives for Germany. 
We conducted the survey in the countries summarised in Table~\ref{tab:datacollection}. The data collection took place in 2014 and 2015, during periods that varied across countries.

\begin{table}[htb]
\small
\centering
\caption{Data collection phases\label{tab:datacollection}}{
\begin{tabular}{lll}
\toprule
\textbf{Area}  & \textbf{Country} & \textbf{Data Collection Phase} \\ \midrule
Central Europe	&	Austria 	&	 2014-05-07 to 2014-09-15		\\
&	Germany 	&	2014-05-07 to 2014-08-18		\\ \addlinespace
North America &	 Canada		&	2014-05-07 to 2015-08-15		\\
& United States of America  	&	2014-05-07 to 2015-05-01		\\ \addlinespace
Northern Europe & Estonia		&	2014-05-07 to 2014-10-31		\\
& Finland		&	2015-06-01 to 2015-08-28		\\
& Norway		&	2014-05-07 to 2014-09-15		\\
& Sweden		&	2014-05-07 to 2014-09-15		\\ \addlinespace
South America 	& Brasil 	&	2014-12-09 to 2015-03-31 \\ \addlinespace		
Western Europe & Ireland  & 2014-05-07 to 2014-12-31 \\  \bottomrule
\end{tabular}}
\end{table}

\subsection{Data Analysis}
\label{sec:DataAnalyis}

The answers to Q~2, which asks for the main business area and application domain, were analyzed qualitatively.
We opted to ask for this information in an open question because there is no well-established standard to
classify companies that are involved in developing software, or to classify application domains.
Moreover, the application domain of a software product may be different from the domain of a company.
For example, an automotive company may only buy its software, develop embedded, safety-critical software or develop primarily its business information systems. 

To reduce subjectivity of interpretation of the qualitative answers to Q~2, 
the answers were coded independently by two different researchers. The interpretations were then compared, and any differences resolved.  We ended up with 50 partly overlapping codes that describe the respondents software and company domains best.

In our preliminary publications on this run of the survey, we compared countries and regions \cite{mendez:softw15,kalinowski:swqd16}. Therefore, we initially also analysed the complete dataset as a whole as well as divided by country and region. Yet, in our first analyses, we found very little differences in the results between countries. Hence, we decided to look into whether the dataset can be considered as coming from one sample.

We conducted a preliminary analysis of the answers collected and checked them with the Kruskal-Wallis (K-W) test~\cite{daniel1990applied}, which determines whether three or more samples originate from the same distribution (in our context, a sample is a country). The K-W test is a non-parametric test, i.e., it does not assume that the data comes from a distribution that can be completely described by two parameters, mean and standard deviation, as in the case of the normal distribution. If the samples come from  the same distribution, the K-W test will show no difference among them\footnote{Unless the populations have symmetrical distributions with the same centre: However, being 10 groups in our case, it is reasonable to assume that it is highly improbable to happen.}, which in our case means that answers to the survey questions are consistent throughout the countries. 

We applied the K-W test with a confidence level $\alpha$ = 0.05: When the p-value is less than 0.05, the hypothesis that the samples originate from the same distribution is rejected. This happens when the answers to a survey question differ significantly depending on the country. We performed the test for each survey question and each corresponding answer option, for a total of 94 tests, as reported in Appendix~\ref{sec:appendix-k-w-tests} (The p-values less than 0.05 are highlighted in bold.).  The null hypothesis had to be rejected in 27 answer options. In 67 tests, the null hypothesis could not be rejected. IF we would correct for multiple testing, this would even be more often the case. In addition, we observe in Tables \ref{tab:kwtestpart1} and \ref{tab:kwtestpart2} in Appendix~\ref{sec:appendix-k-w-tests}, that there is no question in which the null hypothesis was rejected for all its answer options: In many questions, only one or rarely two options for a question lead to rejecting the null hypothesis. 

Therefore, given that (1) in our preliminary publications on this run of the survey, we found very little differences in the results between countries, and (2) the K-W test was rejected only in 1/3 of the answer options, most of them being unrelated, we deemed it reasonable to assume in the theory and the further analyses all requirements engineering practice (related to software projects) world-wide and not country-specific, at least with respect to the countries involved in the survey at present. Including further countries, especially countries with stronger cultural differences, might change that in the future.

This fundamental assumption of our theory, however, does not tell us more about the general population. In such cases of unknown theoretical distribution, resampling methods, and in particular bootstrapping techniques, have been reported to be more reliable and accurate than inference statistic from samples \cite{lunneborg2001bootstrap,ader2008advising}. In addition, non-parametric bootstrap inference is distribution independent (as also non-parametric tests): as a consequence, parametric or sample size assumptions (e.g., assumptions of normality of distribution and of homogeneity of variance) are not necessary as it is not necessary to estimate the mathematical characteristics of the sampling distribution for a sample-based estimate or a test statistic.

We built our analysis on the bootstrapping procedure, i.e.,\ we bootstrap from the sample (we use 1,000 replacements) to statistically infer about the available population \cite{lunneborg2001bootstrap}. In our case, we estimate the means and compute bootstrap 95~\% confidence intervals in the following way:

\begin{itemize}
\item for binary answers (i.e.,\ either the answer is checked or not), we encode a checked option with 1 and 0 otherwise, and then apply bootstrapping to estimate both mean and confidence interval; 
\item for multiple option questions, we consider separately each option, and follow the procedure of binary answers
\item for Likert item questions, we compute the mean on the encoded values for the scale options (from 1 to 5), and then apply bootstrapping to estimate both mean and confidence interval. 
\end{itemize}

It is disputable whether it is reasonable to calculate a mean for Likert items, because they are widely considered to be on an ordinal scale. We agree with this in principle. Yet, it is an issue that has been often debated and using an ``interval assumption permits calculation of means, CIs, and other useful statistics.'' \cite{cumming17} They go on to suggest: ``Researchers very often do calculate means for response data from Likert items. However, before doing so they should think carefully about the strong equal-steps assumption they are making----difficult though that is----and bear in mind that assumption when they report and interpret the mean.'' To support this interpretation, we will also provide median values and provide diagrams with the full distributions. 

Our propositions, at present, are almost all in the form that a technology, activity or reason for applying or not applying a technology or activity exists in the population. Hence, one way to analyse the data would be, if there is a single answer stating the presence of it, we accept the proposition. Yet, the goal of the survey is to build a basis for practice and future research to be able to judge what is \emph{commonly} used in practice. Hence,
we decided to introduce a threshold above which we consider something \emph{commonly used}. To some degree such a threshold will always be arbitrary.
There is no precise understanding of \emph{common}. Another way to look at it is that we would not consider \emph{infrequent} use as \emph{common} use. For this, we can base our quantification on Mosteller and Youtz~\cite{mosteller1990quantifying}, who found that the median value meant when scientists speak of infrequent is 17.3~\%.  We agreed to round this value and that we would consider everything \textbf{above 20~\%} to be in common use. In particular,
we will accept propositions for which the data shows that the confidence interval is 20~\% or higher. In other words, we want to be 95~\% confident that
more than 20~\% of the population would give the corresponding answer. 

Similarly, we handle Likert items. We consider a proposition as supported if the median as well as the lower boundary of the CI are both above 3 (the neutral answer) for positive propositions or below for negative propositions. As we have the neutral answer explicitly, we might also have cases where the median is exactly 3 and the CI also includes it. In that case, we see the corresponding proposition as not supported. Furthermore, as there is also no indication for the opposite of the proposition, we would remove those propositions from the theory for the time being.

Afterwards, we look for explanations for the propositions that we can accept. The explanations either come from (1) existing theories or empirical results from other studies, (2) additional information that we could extract from the answers to open questions in our survey or (3) our own reasoning based on our experiences and discussions among the authors.

\subsection{Validity Procedures}
\label{sec:ValidityProcedures}

The main mechanism to increase the validity of our results were the input and feedback continuously provided by all participating researchers in the joint communities workshops described in section~\ref{sec:NaPiREInitiative}. As for this replication, we started the replication design by a series of Skype calls, each with the main organizers and a regional subset of the other researchers. In those Skype calls, we discussed the theory, potential changes and how the questions for the propositions should look like. We then organized a session with many of the participating researchers and several not involved in the NaPiRE project to review the theory in the 2015 community workshop of the \emph{International Software Engineering Research Network} (ISERN). For this, we brought print-outs of the general theory overview and all the propositions together with potential explanations. We had roughly 20 empirical researchers present who gave written feedback on all aspects they found interesting. This resulted in various minor changes to state the theory more concisely and consistently. 
Furthermore, we used an ``other'' answer option with the possibility to enter free text to capture if we missed an important answer option. We include a discussion of these answers together with the other results.

As for the data collection, we conducted, again, a pilot phase with two practitioners to test the accuracy (and understandability) of the questionnaire, but also the anticipated analysis methods covering both quantitative analyses and qualitative ones.

\section{Detailed Theory and Survey Results}
\label{sec:Results}

In the following, we first summarise the information about the study population, before describing the concrete theory, results and relation to existing evidence for each of the research questions. Since we base our analysis on the same data set that we used in \cite{napire:emse15}, we reuse some of the descriptions and summarise the most important characteristics of our study population. The full data set is openly available \cite{Mendez2018}.

\subsection{Study Population}

In total, 354 organisations spread across 10 countries agreed to answer the survey. Out of these, 228 (63~\%) completed the survey by going through all of its questions and successfully reaching its end (not necessarily answering each question). Table~\ref{tab:population} shows the number of completed questionnaires and the completion rate per country. The completion rates vary mostly between roughly 60~\% and 80~\%. This indicates for us that the completion of the questionnaire is reasonably doable. We do not have an explanation why the completion rate in Sweden was particularly low.

\begin{table}[htb]
\small
\centering
\caption{Study population including response rates, total  obtained, completed datasets, and completion rates. \label{tab:population}}{%
\begin{tabular}{l l r r r r}
\toprule
 &  & \textbf{Response} & \textbf{Total} & \textbf{Completed} & \textbf{Completion}  \\
\textbf{Area}  & \textbf{Country} & \textbf{Rate} & \textbf{Questionnaires} & \textbf{Questionnaires} & \textbf{Rate}  \\ \midrule
Central Europe	&	Austria	&	72.0~\% & 18	&	14	&	78~\% 	\\
&	Germany	&	36.8~\% & 50	&	41	&	82~\%  \\ \addlinespace

North America	&	Canada	&	75.0~\% & 15	&	13	&	87~\% \\
&	USA	&	36.2~\% & 25	&	15	&	60~\% \\ \addlinespace
Northern Europe	&	Estonia	&	25.7~\% &9	&	8	&	89~\%	\\
&	Finland	&	37.5~\% &	18 &	15	&	83~\% \\
&	Norway	&	70.8~\% &	17 &	10	&	59~\% \\
&	Sweden	&	 51.8~\% & 59	&	20	&	34~\% \\ \addlinespace
South America	&	Brazil	&	35.3~\% & 118	&	74	&	63~\% \\ \addlinespace
Western Europe &	Ireland	&	39.7~\% & 25	&	18	&	72~\%  \\\midrule
Total &  &  & 354 & 228 & 64~\% \\
\bottomrule
\end{tabular}}
\end{table}

\begin{figure}[!hbtp]
\centering
  \includegraphics[width=0.8\textwidth]{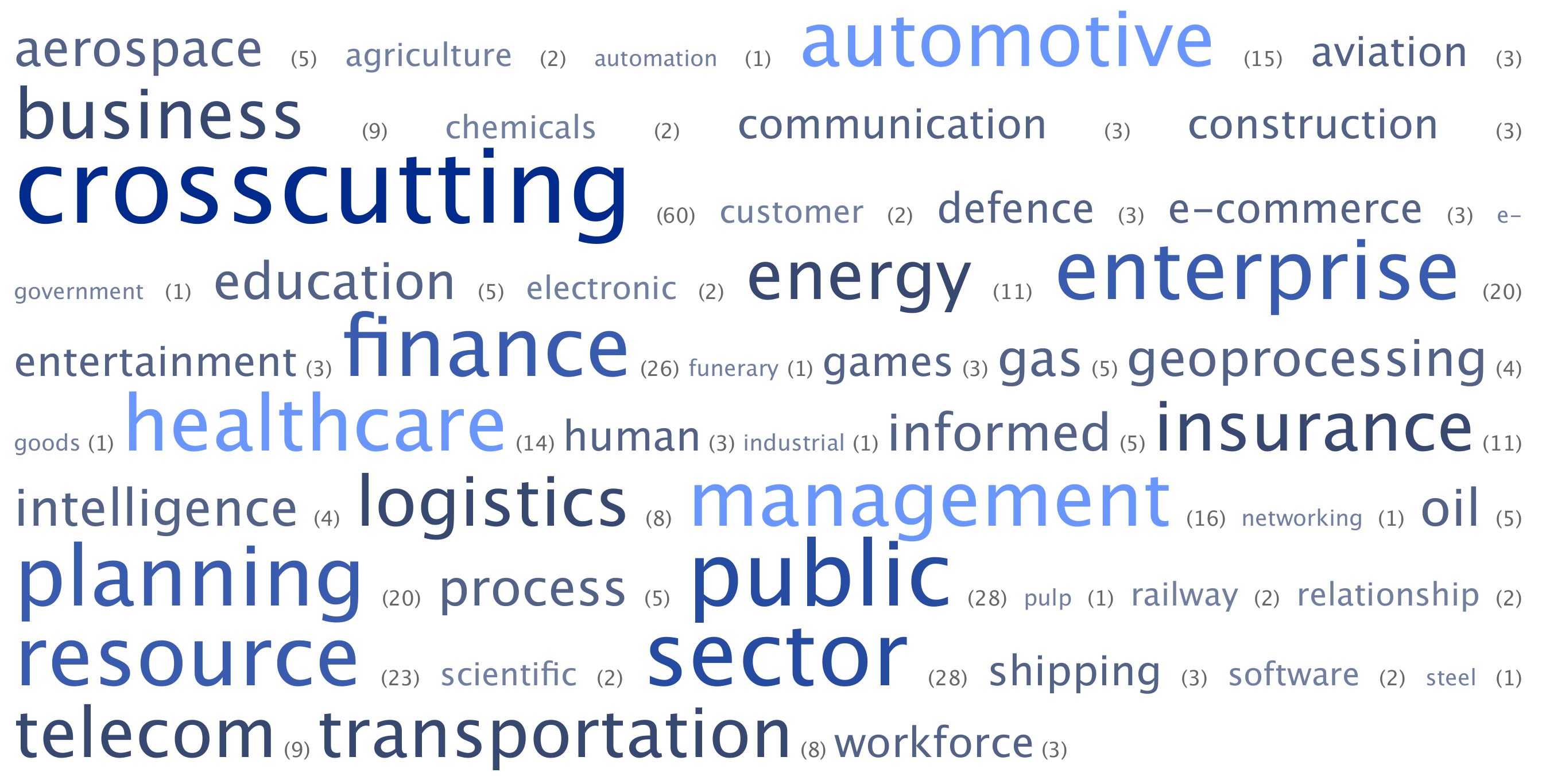}\\
  \caption{Tag cloud of the business domains of the responding organisations.}
  \label{fig:BusinessDomains}
\end{figure}

The results reported in this article consider the completed datasets only. 
The domains were provided by the respondents in free text format (see~Table~\ref{tab:instrument}, question Q2) and coded by the researchers. The tag cloud for the coded business domains can be seen in Figure~\ref{fig:BusinessDomains}.

This figure shows the frequency of each domain code and highlights the most frequent ones. Of the 228 organisations, 215 provided answers for their business domain. We found a huge variety in business domains, ranging from embedded software systems to consulting. Many organisations were actively working with products and/or services that can be used in several business domains (e.g.\ cloud computing and web applications, custom software development, enterprise resource planning products, IT consulting services).

Additionally, we identified a very large number of different business areas and application domains with few data points in each one. Therefore, considering the number of organisations active in several business areas and application domains and the large variety of different areas and domains reported, we decided to characterise the responding organisations independent of their business areas and application domain, i.e.,\ in terms of the characteristics 'size' and 'process model used' (see also Sect.~\ref{sec:DataAnalyis}) in Table~\ref{tab:organisationsizes}. Concerning size, we grouped organisations as small, medium, and large-sized according to the number of employees (software and other areas). 

\begin{table}[htb]
\scriptsize
\centering
\caption{Sizes of responding organisations.\label{tab:organisationsizes}}{%
\begin{tabular}{lrrrrrrr}
\toprule
& \textbf{Central} & \textbf{North} & \textbf{Northern} & \textbf{South} & \textbf{Western}\\
\textbf{Size}  & \textbf{Europe} & \textbf{America} & \textbf{Europe} & \textbf{America} & \textbf{Europe} & \textbf{Total} \\ \midrule
Small ($\leq$ 50 employees)& 4 & 11 & 20 & 26 & 8 & 69 \\
Medium (51 to 250 employees)& 1 & 0 & 12 & 17 & 3 & 33 \\
Large ($\geq$ 251 employees)& 29 & 16 & 34 & 28 & 7 & 114 \\ \midrule
Total & 34 & 27 & 66 & 71 & 18 & 216 \\
\bottomrule
\end{tabular}}
\end{table}

We can observe that the datasets include relatively large samples of both, small and large-sized organisations. Considering the distributions of size per region, except for South-America, the responding large-sized organisations slightly outweigh the small and medium-sized organisations. Note that not all respondents answered this question.

Regarding the process models used, respondents answered a multiple choice question with the following options: RUP \cite{kruchten2003}, Scrum, V-Model XT, Waterfall, XP, and Other (in this case informing us textually which process model they use). We grouped these process models into \emph{agile} (Scrum and XP), \emph{plan-driven} (RUP, Waterfall and V-Model XT), and \emph{mixed} (for those organisations that indicated that they use agile and plan-driven process models or variations therein). Out of the 228 organisations that completed the questionnaire, 196 selected one of the five predefined options for their process model. 
See Table~\ref{tab:swprocessmodels}.

\begin{table}[htb]
\scriptsize
\centering
\caption{Software process models used in responding organisations.\label{tab:swprocessmodels}}{%
\begin{tabular}{lrrrrrrr}
\toprule
& \textbf{Central} & \textbf{North} & \textbf{Northern} & \textbf{South} & \textbf{Western}\\
\textbf{Model}  & \textbf{Europe} & \textbf{America} & \textbf{Europe} & \textbf{America} & \textbf{Europe} & \textbf{Total} \\ \midrule
Agile & 4 & 13 & 35 & 32 & 8 & 92 \\
Plan-driven & 13 & 4 & 8 & 19 & 2 & 46 \\
Mixed & 12 & 8 & 19 & 14 & 5 & 58 \\ \midrule
Total & 29 & 25 & 62 & 65 & 15 & 196 \\
\bottomrule
\end{tabular}}
\end{table}

Again, the dataset includes relatively large samples of both, agile and plan-driven organisations. Considering the distribution per region, except for Central Europe, the responding organisations following agile process models in the respondents environment outweigh the plan-driven ones. The number of organisations using mixed process models (or more than one) is large. 

We therefore could obtain a balanced characterisation of small, medium and large organisations of different regions using both, plan-driven and agile development methods.

\subsection{Status Quo in Requirements Elicitation and Documentation}

One of the core activities in requirements engineering is eliciting the requirements from relevant stakeholders.
To characterise the status quo, we want to understand what elicitation techniques are employed in practice.
In our theory from the first run, we expected practitioners, especially in large companies, to conduct workshops as the
central technique to elicit requirements. The first run, however, showed that other elicitation techniques
are also widely in use~\cite{MW14}. Therefore, we widened the choice of elicitation techniques as shown
in Table~\ref{tab:p-elicitation}. To make it consistent with common terminology, we adopted the elicitation techniques as described in the SWEBoK~\cite{bourque2014guide}. Table~\ref{tab:p-elicitation} also notes whether the corresponding proposition was supported in the first run or if it is a new proposition for this run.

\begin{table}[!htb]
\centering \scriptsize
\caption{Propositions about the status quo in requirements engineering elicitation prior to the survey \label{tab:p-elicitation}}{%
\begin{tabular}{lp{0.55\linewidth}ll}
\toprule
 & & \textbf{Supported in} & \textbf{Survey}\\
\textbf{No.} & \textbf{Propositions} & \textbf{first run or new} & \textbf{question}\\\midrule
P~1 & Requirements are elicited via interviews. & New & Q~9 \\
P~2 & Requirements are elicited via scenarios. & New & Q~9 \\
P~3 & Requirements are elicited via prototyping. & New & Q~9 \\
P~4 & Requirements are elicited via facilitated meetings (including workshops). & Supported & Q~9 \\
P~5 & Requirements are elicited via observation. & New & Q~9 \\ \bottomrule
\end{tabular}}
\end{table}

\begin{figure}[!htb]
\centering
\includegraphics[width=120mm]{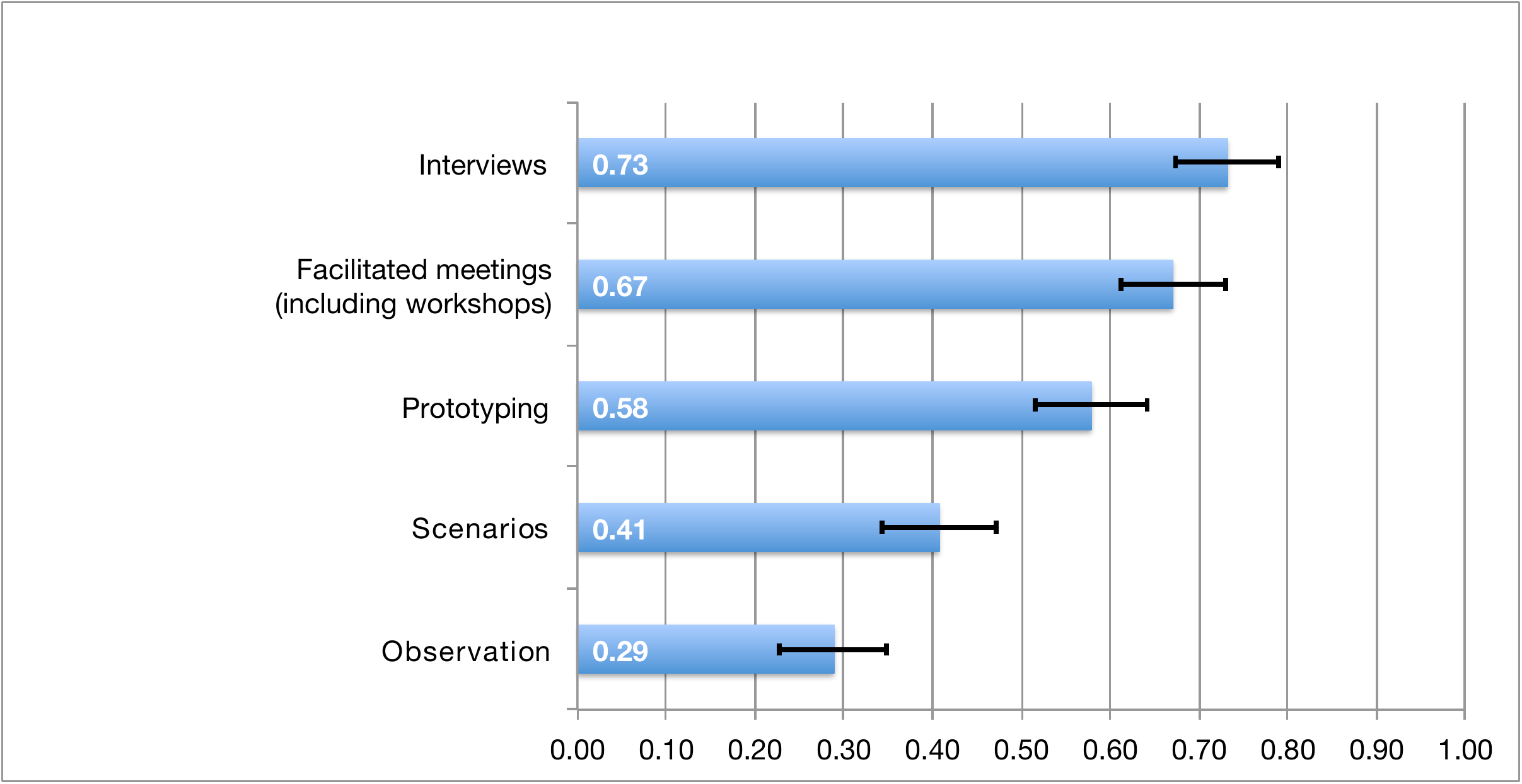}
\caption{How do you elicit requirements? ($N = 228$)}\label{fig:elicit}
\end{figure}

The answer possibilities in the questionnaire correspond directly to the propositions. The proportion of how often the elicitation techniques from the propositions have been selected by our respondents is shown in Fig.~\ref{fig:elicit} together with an error bar that represents the 95~\% confidence interval (CI). The $N$ in the caption denotes the number of participants that answered this question. We will report the proportion $P$ of the participants that checked the corresponding answer and its 95~\% confidence interval in square brackets in the following. The most frequently used techniques are interviews with $P = 0.73$ [0.67, 0.79] and facilitated meetings with $P = 0.67$ [0.61, 0.73] closely followed by prototyping ($P = 0.58$ [0.52, 0.64]) and scenarios ($P = 0.41$ [0.34, 0.47]). Observations reached only a $P = 0.29$ [0.23, 0.35]. Yet, it is still larger than the threshold of 0.2. 

Additional answers for ``others'' included ``Created personas and presented them to our stakeholders'', ``Questionnaires''/``Surveys'' , ``Analysis of existing system'' and ``It depends on the client.'' Especially some kind of surveys/questionnaires are mentioned several times. This could be a candidate for an additional
proposition for the next theory.

We therefore have support for all corresponding propositions P1 to P5. All five mentioned elicitation techniques are used in practice. For interviews, facilitated meeting and prototyping, each is used by more than half of the respondents. Scenarios and observations are less often used but are still not niche techniques.

Comparing the confidence intervals, we can also generalise that interviews, facilitated meetings and prototyping are the three top techniques. Their intervals overlap so that we cannot distinguish them in general. They are, however, significantly more used than scenarios and observations which again overlap.

For explanations of the propositions, we do not have any additional insights from the open answers. It is often stated in the literature that it is
important to include different viewpoints during requirements engineering (supporting early work by Sommerville \textit{et al}.~\cite{667811}). Interviews and facilitated meetings are probably the easiest ways to collect many viewpoints. Prototyping and scenarios are ways to represent the system. There it depends on the number of people a requirements engineer shows them, how many viewpoints they get. The value of prototypes and scenarios to trigger requirements from different stakeholders (including non-technical end users) has been reiterated by Mannio and Nikula~\cite{MN01} who developed an iterative requirements elicitation method combining these two approaches. Observations, finally, are probably often the most difficult and time-consuming way to get an understanding of different viewpoints. We formulate the viewpoint aspect as explanation E~1 in Tab.~\ref{tab:e-elicitation}. Furthermore, we include the theory of Mannio and Nikula~\cite{MN01} as E~2 that prototypes and scenarios are helpful for a shared understanding of the requirements among stakeholders. Overall, Tab.~\ref{tab:e-elicitation} shows that we could leave all propositions unchanged and summarise the explanations.

\begin{table}[htb]
\centering \scriptsize
\caption{Propositions about elicitation with explanations after the survey\label{tab:e-elicitation}}{%
\begin{tabular}{lp{0.7\linewidth}l}
\toprule
\textbf{No.} & \textbf{Propositions} & \textbf{Changed} \\\midrule
P~1 & Requirements are elicited via interviews.  \\
P~2 & Requirements are elicited via scenarios.  \\
P~3 & Requirements are elicited via prototyping.  \\
P~4 & Requirements are elicited via facilitated meetings (including workshops).  \\
P~5 & Requirements are elicited via observation.  \\ \midrule
\textbf{No.} & \textbf{Explanations} & \textbf{Propositions}  \\\midrule
E~1 & Interviews, scenarios, prototyping, facilitated meetings and observations allow the requirements engineers to include many different viewpoints including those from non-technical stakeholders & P1--P5 \\ 
E~2 & Prototypes and scenarios promote a shared understanding of the requirements among stakeholders & P2, P3\\\bottomrule
\end{tabular}}
\end{table}


The second major activity is the documentation of the elicited requirements. Here,
we stated propositions along two dimensions: (1) the type of document such as structured
requirements lists or use case models and (2) the level of formality. 
Possible requirements document types are \emph{structured requirements lists}, \emph{use case models}, \emph{domain/business
process models}, \emph{goal models} and \emph{data models} because they are often mentioned
in practice and/or research. The level of formality is either \emph{textual free form with no constraints}, \emph{textual with constraints} such as the user story template (``As a\ldots, I want to\ldots, so that\ldots''),
\emph{semi-formal} such as UML diagrams or \emph{formal}
with a mathematical basis and formal semantics. Furthermore, we briefly go into non-functional requirements and expect them to be documented in a non-quantified and textual way. The propositions of our theory related to
requirements documentation are given in Table~\ref{tab:p-documentation}. They are all new
in relation to the theory from the first run.

\begin{table}[!htb]
\centering \scriptsize
\caption{Propositions about the status quo in requirements engineering documentation prior to the survey\label{tab:p-documentation}}{%
\begin{tabular}{lp{0.6\linewidth}ll}
\toprule
 & & \textbf{Supported in} & \textbf{Survey}\\
\textbf{No.} & \textbf{Propositions} & \textbf{first run or new} & \textbf{question}\\\midrule
P~6 & Structured requirements lists are documented textually in free form. & New & Q~10 \\
P~7 & Use case models are documented textually in free form. & New & Q~10 \\
P~8 & Use case models are documented semi-formally (e.g.\ using UML). & New & Q~10 \\
P~9 & Domain/business process models are documented semi-formally (e.g.\ using UML). & New & Q~10 \\
P~10 & Goal models are documented semi-formally (e.g.\ using UML). & New & Q~10 \\
P~11 & Goal models are documented formally. & New & Q~10 \\
P~12 & Data models are documented semi-formally (e.g.\ using UML). & New & Q~10\\
P~13 & Non-functional requirements are documented in a non-quantified and textual way. & New & Q~11\\  \bottomrule
\end{tabular}}
\end{table}

\begin{figure}
\includegraphics[width=\textwidth]{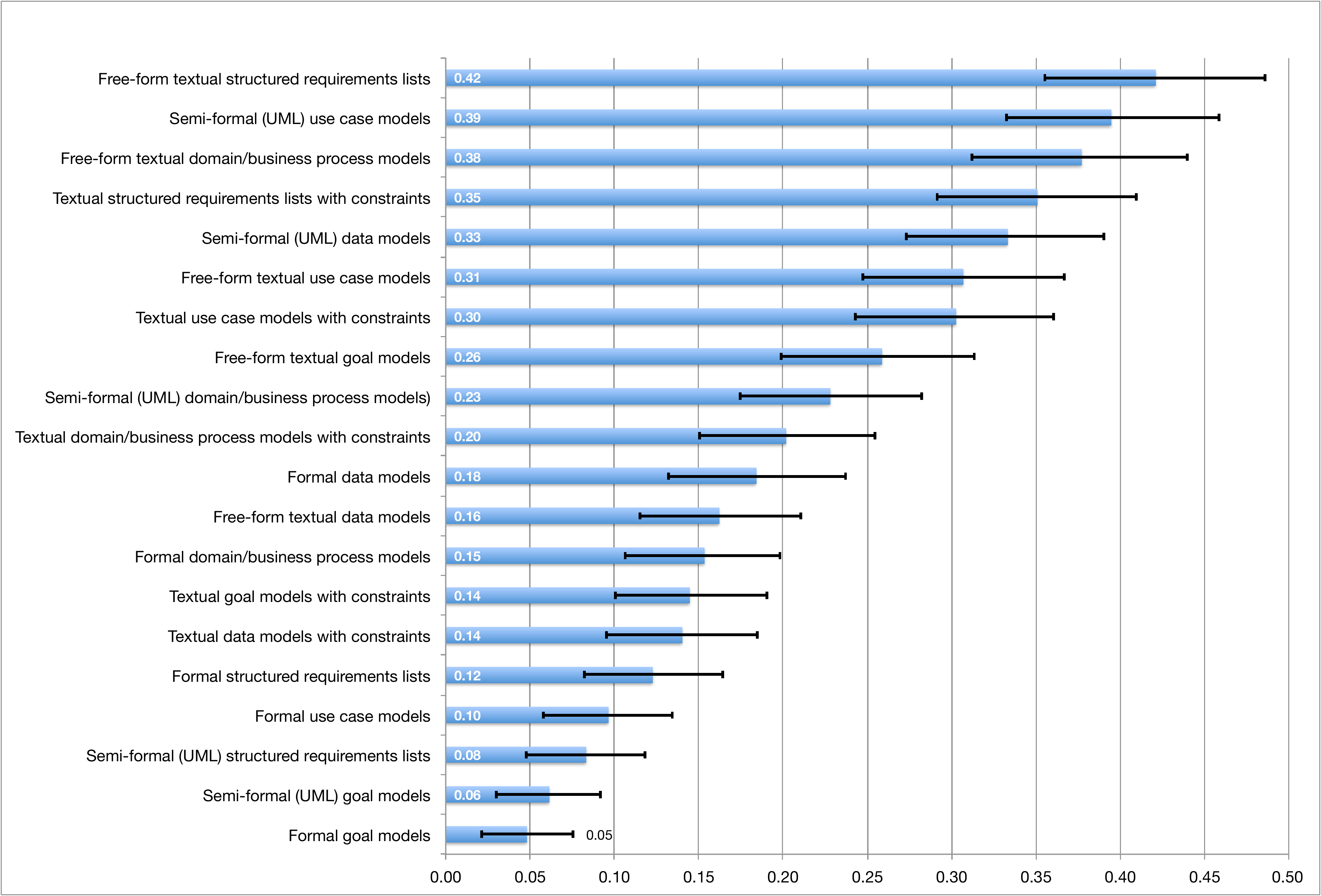}
\caption{How do you document functional requirements? ($N = 228$)}\label{fig:document-functional}
\end{figure}

In the questionnaire, the
respondents could choose multiple items from the description techniques with a degree of formality. As shown in Fig.~\ref{fig:document-functional}, the three most frequent ways to document requirements are free-form textual structured requirements lists ($P = 0.42$ [0.36, 0.49]), semi-formal use case models ($P = 0.39$ [0.33, 0.46]) and free-form textual domain/business process models ($P = 0.38$ [0.31, 0.44]). But also textual requirements lists with constraints -- e.g., user stories -- ($P = 0.35$ [0.29, 0.41]), semi-formal data models ($P = 0.33$ [0.28, 0.39]), free-form textual ($P = 0.31$ [0.25, 0.37]) as well as constraint textual use cases ($P = 0.30$ [0.24, 0.36]) were mentioned often. Free form textual goal models ($P = 0.26$ [0.20, 0.32]) are
very close to the threshold. None of the documentation techniques seems to be clearly dominating.

All other documentation techniques fall below our threshold by including 0.20 in their CI.  Semi-formal domain/business models ($P = 0.23$ [0.17, 0.28]), domain/business model with constraint text
($P = 0.20$ [0.15, 0.25]), formal data models ($P = 0.18$ [0.13, 0.24]) and free-form textual data models ($P = 0.16$ [0.11, 0.21]) all are around 20~\%. The remaining techniques are then clearly below 20~\%: Formal business/domain models have $P = 0.15$ [0.11, 0.20], textual goal models
with constraints have $P = 0.14$ [0.10, 0.19], textual data models with constraints have $P = 0.14$ [0.10, 0.19] and formal structured requirements
lists have $P = 0.12$ [0.08, 0.16]. Rarely used are formal use case models ($P = 0.10$ [0.06, 0.13]), semi-formal structured requirements lists
($P = 0.08$ [0.05, 0.12] as well as semi-formal ($P = 0.06$ [0.03, 0.09]) and formal goal models ($P = 0.05$ [0.02, 0.07]).

Proposition P~6, which states that structured requirements lists are documented textually in free form, is clearly supported by our data. Yet, we see textual requirements with constraints on the same level. Hence, we will update our proposition to also include this kind of requirements documentation. An explanation for this could be that there are many requirements where text is sufficient be it free-form or constrained. Especially in agile projects and user stories, more than text is often not needed, because ``the main purpose of a story card is to act as a reminder to discuss the feature.''~\cite{cohn2004} We use this as explanation E~3 in Table~\ref{tab:e-documentation}.

\begin{table}[!htb]
\centering \scriptsize
\caption{Propositions about requirements documentation and explanations after the survey\label{tab:e-documentation}}{%
\begin{tabular}{lp{0.7\linewidth}l}
\toprule
\textbf{No.} & \textbf{Propositions} & \textbf{Changed} \\\midrule
P~6 & Structured requirements lists are documented textually in free form or textually with constraints. & \checkmark \\
P~7 & Use case models are documented textually in free form or textually with constraints. & \checkmark \\
P~8 & Use case models are documented semi-formally (e.g.\ using UML). &  \\
P~9 & Domain/business process models are documented textually in free form. & \checkmark \\
P~10 & Goal models are commonly used in a textual form. & \checkmark \\
P~11 & Goal models are not documented semi-formally or formally. & \checkmark \\
P~12 & Data models are documented semi-formally (e.g.\ using UML). &  \\
P~13 & Non-functional requirements are documented textually either quantified or non-quantified. & \checkmark \\  \midrule
\textbf{No.} & \textbf{Explanations} & \textbf{Propositions} \\\midrule
E~3 & Free-form and constraint textual requirements are sufficient for many contexts such as in agile projects where they only act as reminders for further conversations. & P~6, P~7, P~9--11\\
E~4 & Use case models and data models might not often be shared with non-technical stakeholders. Hence, requirements engineers can use well-known semi-formal description techniques such as entity-relationship diagrams or UML to document them. & P~8, P~12\\
E~5 & The quantification depends on the type of non-functional requirement. Performance is rather documented quantitatively while maintainability is rather documented non-quantitatively. & P~13\\
 \bottomrule
\end{tabular}}
\end{table}

We also have good support for propositions P~7 and P~8. Documentation in the form of semi-formal use case models was the second most popular answer. Also free-form textual descriptions of use cases were mentioned by 31~\% of the respondents. Yet, we again have the textual documentation with constraints on a similar level. Hence, we will extend P~7 to include this kind of documentation. Formal use cases, in contrast, are significantly less used. For the textual description, we can use E~3 again as explanation. For the semi-formal description, we believe that those documents might not be shared with non-technical stakeholders. Then the engineers might tend to use UML which is now widely taught in degree programs (E~4).

Proposition P~9 stated that domain/business process models are documented semi-formally, e.g.\ using UML or some other standardized notation such as BPMN. This is not well supported in our data. The CI goes below the 20~\% threshold. In contrast, the most often mentioned way to document domain or business process models was free-form textual with 38~\%. Hence, we will replace P~9 with ``Domain and business process models are documented textually in free form.'' It seems that E~3 also holds for domain and business process models.

Both propositions P~10 and P~11 have no support in the data. They state that goal models are either formally or semi-formally documented. However, these two options were chosen by very few of the respondents. Both CI are below 20~\%. Instead, goal models are actually most often documented in a free-form textual way (26~\% of the respondents). The CI only touches our 20~\% threshold. The confidence interval is not overlapping with the intervals of formal and semi-formal documentation. Therefore, we will replace these propositions with two new proposition on goal models: ``Goal models are commonly used in a textual form.'' and ``Goal models are not documented semi-formally or formally.'' An explanation for these new proposition could be again E~3. 

Proposition P~12 states that data models are documented semi-formally. This is well supported by the data. The corresponding
answer is the most frequently chosen one for data models. Also considering the confidence intervals it is beyond the threshold. Furthermore, the semi-formal way of documentation is used significantly more often then the other ways to document data models. Data models are probably not often directly discussed with (non-technical) customers and users and, hence, requirements engineers can use well-known semi-formal techniques such as entity-relationship diagrams or corresponding UML class models. This is E~4 in Table~\ref{tab:e-documentation}.

\begin{figure}[!htb]
\centering\includegraphics[width=\textwidth]{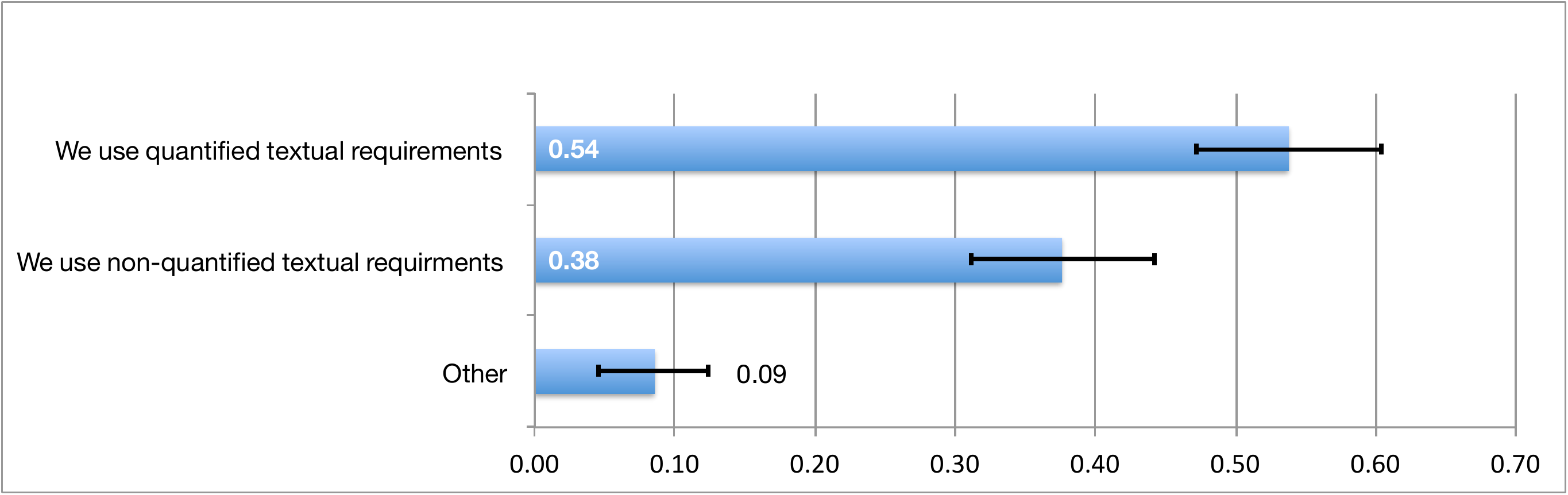}
\caption{How do you document non-functional requirements? ($N = 210$)}\label{fig:document-non-functional}
\end{figure}

Finally for RQ~1, we briefly touched upon the topic of non-functional requirements (such as security or performance requirements).
As shown in Fig.~\ref{fig:document-non-functional}, we found that about half of the respondents (P = 0.54 [0.47, 0.60]) use quantified textual documentation for non-functional requirements. Only about 38\% state that they document non-functional requirements not in a quantified way ($P = 0.38$ [0.31, 0.44]).  Answers for \emph{Other} included ``on our user stories'', ``user story acceptance criteria'' or ``both of the above''. It was
chosen with $P = 0.09$ [0.05, 0.13].

Therefore, we have support for proposition P~13 that non-functional requirements are documented in a non-quantified
and textual way. Yet, most respondents seem to quantify their non-functional requirements. Even the confidence intervals do not overlap. We therefore accept proposition P~13 but replace it with ``Non-functional requirements are documented textually either quantified or non-quantified.'' An explanation could be that the respondents primarily thought about performance requirements that are often quantified (E~5). Yet, other non-functional requirements such as security or maintainability are much harder to quantify. In further studies, we need to distinguish here more clearly.

Cox, Niazi and Verner~\cite{CNV09} found that for documentation, making a business case of a project was most valuable. This adds probably a more general dimension to requirements documentation that we have not covered. Similarly, for elicitation, they found assessing system feasibility to be most valuable. Our results fit together in that they found that valuable techniques are to specify requirements quantitatively. 
We can also support Nikula, Sajaniemi and K\"alvi\"ainen~\citeN{nikula2000sps} that requirements documentation is often done textually (natural language).

We have some contradicting results to Neil and Laplante~\citeN{Neill:2003ui}. They found that most respondents use scenarios \& use cases and many use focus groups and informal modelling, while interviews and prototypes are far less used. In contrast, we saw scenarios as significantly less used than interviews, facilitated meetings and prototyping. A reason might be that scenarios and use cases were very popular when Neil and Laplante did their study, and the popularity might have declined over the last 15 years. Yet, we can support their findings that most requirements documentations are informal or semi-formal. 

Their update in 2014 \citeN{Kassab2014} came much closer to our results. Now, they also found interviews and prototyping to be widely used techniques, while scenarios were not as dominant as in their earlier study. A contradiction is still that workshops are only at a bit more than 20~\% while 67~\% of our respondents mentioned them. A reason might be that \citeN{Kassab2014} included further detailed elicitation techniques that could also be seen as facilitated meetings, such as card sorting or group work. 

In guidelines derived from a systematic literature review of empirical studies on elicitation techniques, Dieste and Juristo~\cite{dieste2011systematic} state that especially unstructured interviews are more effective and output more complete information than introspective techniques or sorting techniques.

Marvin et al.~\cite{marvin17} asked requirements engineering practitioners with a questionnaire and found ``that use of goals in practice is inconsistent, informal, and rarely utilises formal modelling approaches.'' This is fully in line with our results of goals only being used informally.

\subsection{Status Quo in Requirements Engineering Changes and Alignment}

Requirements have to be continuously updated to guarantee project success and customer satisfaction. For the state of practice in change management, we therefore expected already in the initial study that a requirements change management is established after a requirements specification (expected to be complete) was formally accepted. Due to the importance of requirements changes, we added several propositions, which are shown in Table~\ref{tab:p-changes}. With regard to requirements changes, we added propositions that product backlogs are updated due to requirements changes after the initial release as well as that requirements changes after the initial release are reflected only in change requests. With regard to traceability, we added propositions that trace between requirements and code as well as between requirements and design documents are explicitly managed. Furthermore, we add propositions that for analysing the effect of requirements changes, impact analysis on code is done, but impact analysis between requirements is not performed.

\begin{table}[!htb]
\centering \scriptsize
\caption{Propositions about the status quo in requirements changes before the survey \label{tab:p-changes}}{%
\begin{tabular}{lp{0.6\linewidth}ll}
\toprule
 & & \textbf{Supported in} & \textbf{Survey}\\
\textbf{No.} & \textbf{Propositions} & \textbf{first run or new} & \textbf{question}\\\midrule
P~14 & A requirements change management is established after formally accepting a requirements specification. & Supported & Q~21\\
P~15 & Product backlogs are updated because of requirements changes after the initial release. & New & Q~12\\
P~16 & Requirements changes after the initial release are reflected only in change requests.& New & Q~12\\
P~17 & Traces between requirements and code are explicitly managed. & New & Q~13\\
P~18 & Traces between requirements and design documents are explicitly managed. & New & Q~13\\
P~19 & For analyzing the effect of changes to requirements, impact analysis on the code is done. & New & Q~14\\
P~20 & For analyzing the effect of changes to requirements, impact analysis between requirements is not done. & New & Q~14\\
\bottomrule
\end{tabular}}
\end{table}

\begin{figure}[!htb]
	\includegraphics[width=\textwidth]{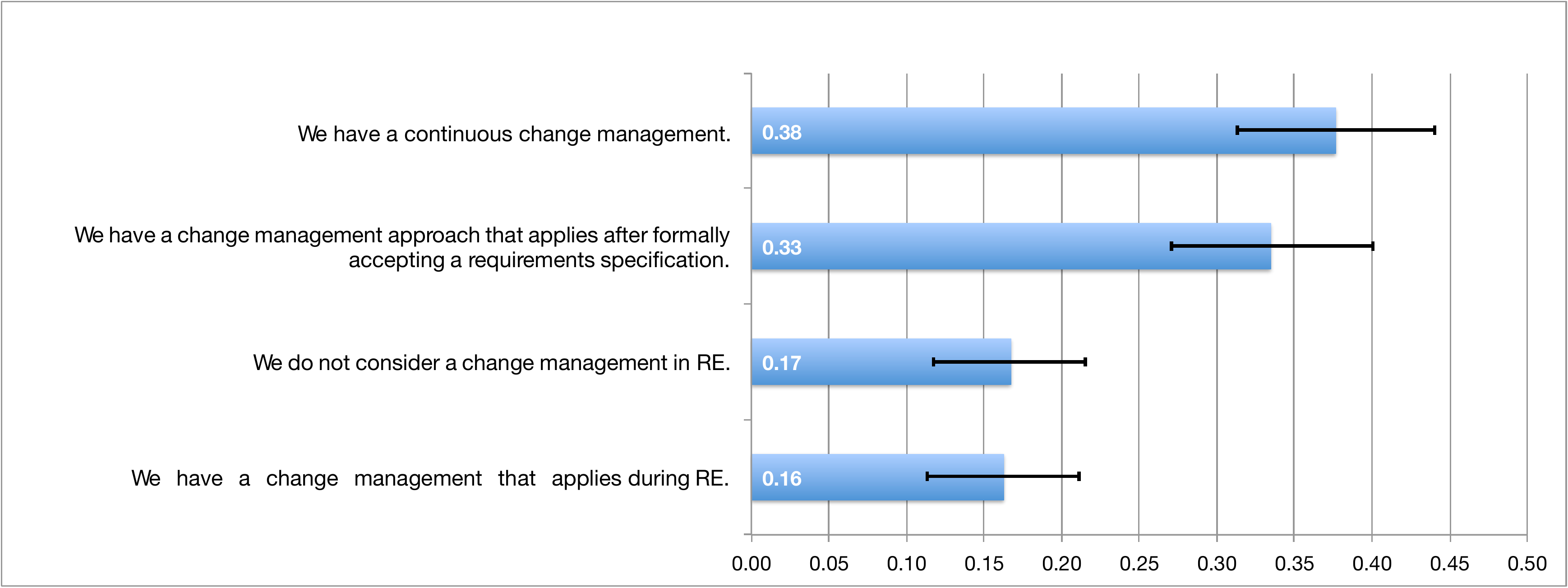}
	\caption{How do you perform change management in your requirements engineering? ($N = 215$)}\label{fig:re-change-management}
\end{figure}

In the survey, we first asked how the respondents perform change management in their requirements engineering process. As shown in Fig.~\ref{fig:re-change-management}, most respondents have a continuous change management ($P = 0.38$ [0.31, 0.44]) or a change management approach that applies after formally accepting a requirements specification ($P = 0.33$ [0.27, 0.40]). Not considering change management in RE ($P = 0.17$ [0.11, 0.21]) or not having change management during RE ($P = 0.16$ [0.12, 0.22]) are less frequently applied. The latter two options are also considering the confidence intervals significantly less used.

Proposition P~14 stating that requirements change management is established after formally accepting a requirements specification is supported by the survey data. P~14 can be explained by E~6 in Table~\ref{tab:e-changes}. Given the high proportion and CI of continuous change management, 
we will add a further proposition: ``Organisations use continuous change management.'' This new proposition could be explained by the continuous 
nature of change in agile development processes. Both other answer options lie below the threshold although both CI include it. 

\begin{table}[!htb]
\centering \scriptsize
\caption{Propositions about requirements changes and explanations after the survey (Q~12, Q~13, Q~14, Q~21)\label{tab:e-changes}}{%
\begin{tabular}{lp{0.7\linewidth}l}
\toprule
\textbf{No.} & \textbf{Propositions} & \textbf{Changed} \\\midrule
P~14 & A requirements change management is established after formally accepting a requirements specification. & \\
P~14a & Organisations use continuous change management. & \checkmark \\
P~15 & Product backlogs are updated because of requirements changes after the initial release. &  \\
P~16 & Requirements changes after the initial release are reflected only in change requests.&  \\
P~17 & Traces between requirements and code are explicitly managed. &  \\
P~18 & Traces between requirements and design documents are explicitly managed. &  \\
P~19 & For analyzing the effect of changes to requirements, impact analysis on the code is done. &  \\
P~20 & For analyzing the effect of changes to requirements, impact analysis between requirements is not done. & \\
\midrule
\textbf{No.} & \textbf{Explanations} & \textbf{Propositions} \\\midrule
E~6 & In many development processes, requirements are fixed at some point(s) in time. A formal change management is only needed afterwards. & P~14 \\
E~7 & In agile development process, change is continuous. & P~14a\\
E~8 & Requirements change during a development project and also after the initial release. Many organisations only work with change requests in issue trackers. Agile organisations work with some kind of product backlog (as in Scrum) and change it regularly between iterations. & P~15, P~16 \\
E~9 & Explicit traces make impact analysis more effective and efficient.  & P~17, P~18 \\	
E~10 & Despite traces between requirements and code, the effect of changes is most directly seen on the code level. & P~19, P~20 \\
\bottomrule
\end{tabular}}
\end{table}

\begin{figure}[!htb]
	\centering\includegraphics[width=\textwidth]{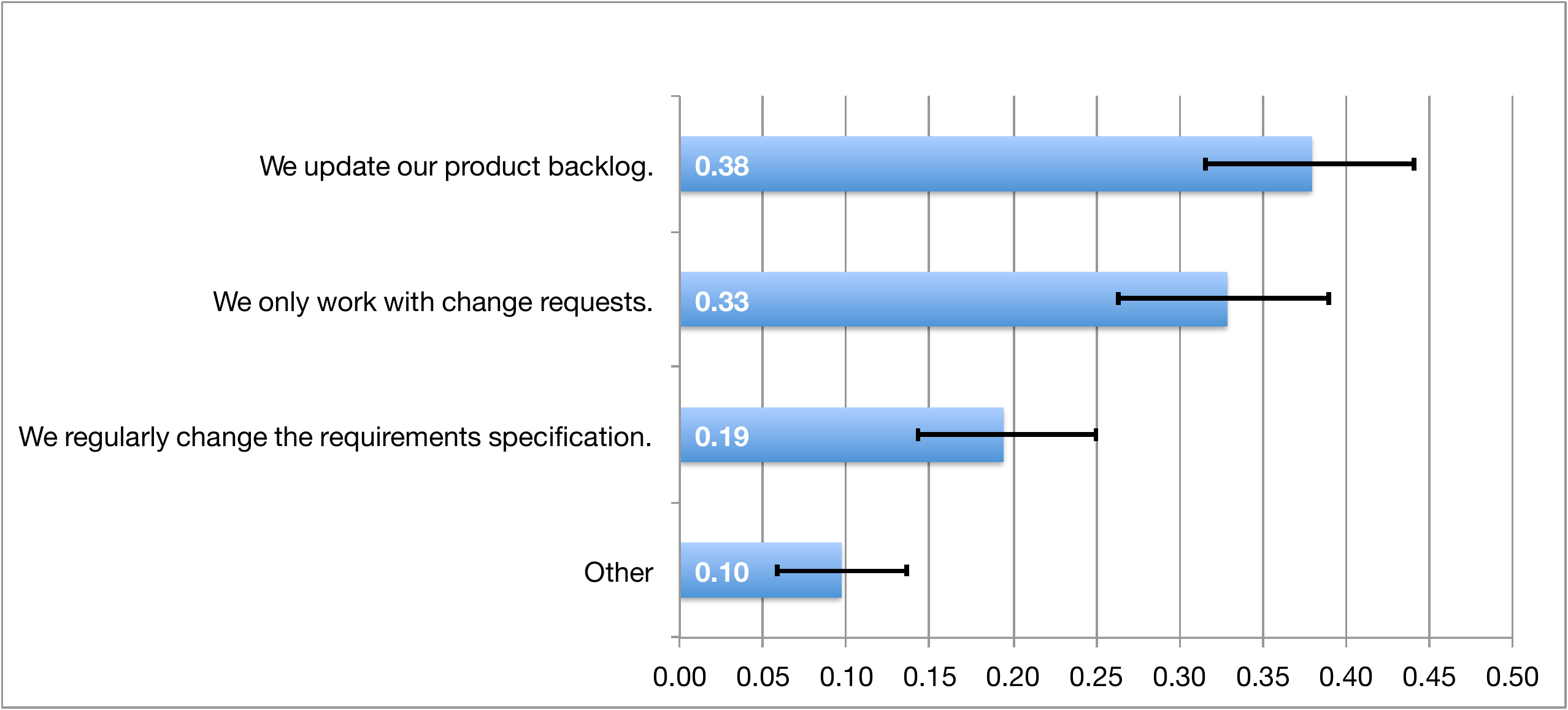}
	\caption{How do you deal with changing requirements after the initial release? ($N = 216$)}\label{fig:deal-with-change}
\end{figure}

Second, we asked how the respondents deal with changing requirements after the initial release. The answers are shown in Fig.~\ref{fig:deal-with-change}. The most common way to do so is to update the product backlog ($P = 0.38$ [0.32, 0.44]). But also working only with change requests is very common ($P = 0.33$ [0.27, 0.39]). Regular changes in the requirements specification are much less used ($P = 0.19$ [0.14, 0.25]). In fact, the confidence interval of the latter does not overlap with the two most popular answers. Answers for \emph{Other} ($P = 0.10$ [0.06, 0.14]) include ``all methods, depends on the project'' and ``we mix product backlog and change requests.'' 

Hence, proposition P~15 on the product backlog update is supported by the data. Proposition P~16 on requirements change only done by change requests is similarly well supported. Propositions P~15 and P~16 can be explained by E~8.

\begin{figure}[!htb]
\centering\includegraphics[width=\textwidth]{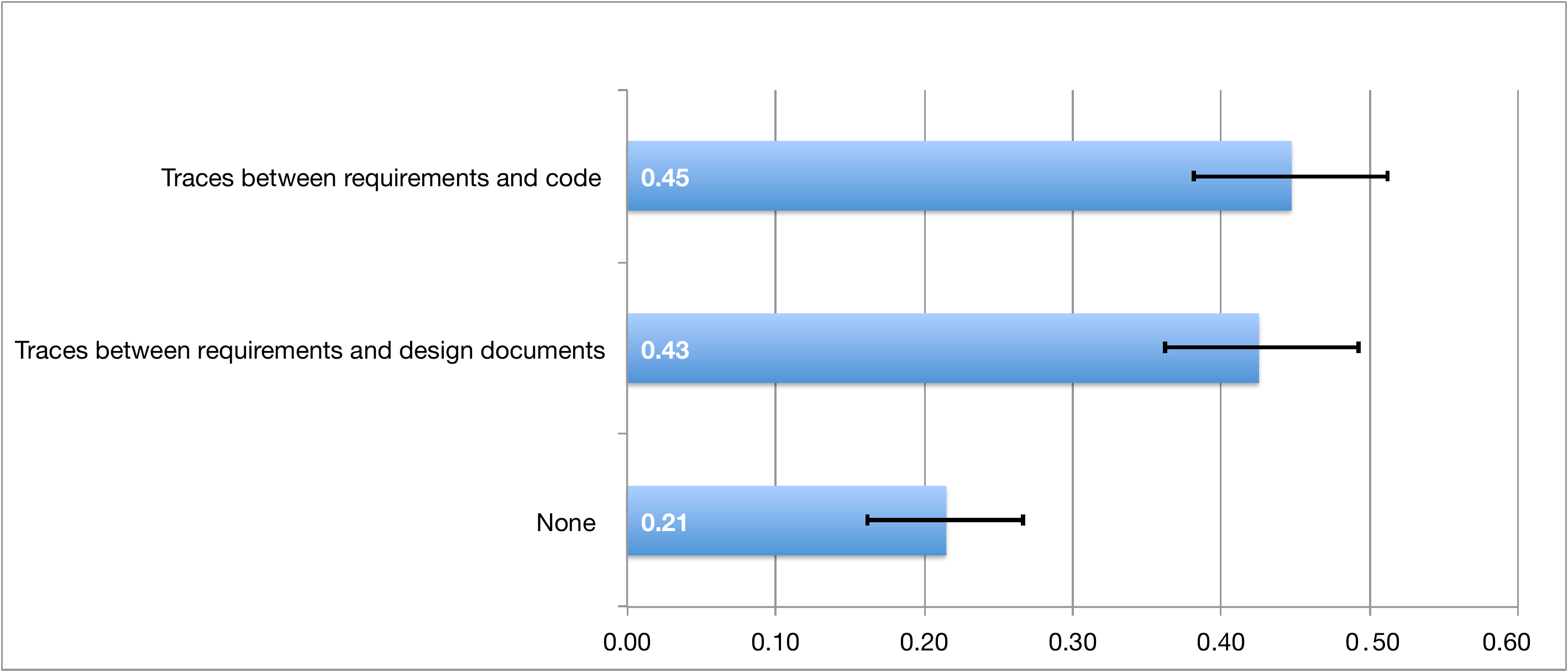}
\caption{Which traces do you explicitly manage? ($N = 228$)}\label{fig:traces}
\end{figure}

As can be seen in Fig.~\ref{fig:traces}, traces between requirements and code ($P = 0.45$ [0.38, 0.51]) as well as between requirements and design documents ($P = 0.43$ [0.36, 0.49]) are often explicitly managed. It is not common that traces are not managed at all ($P = 0.21$ [0.16, 0,27]). For the \emph{Other} answer, several respondents mentioned traces between tests and requirements: ``Traces between requirements and functional/system tests are most common for us.'' Propositions P~17 and P~18 state that traces between requirements and code and between requirements and design documents are explicitly managed. The data supports these propositions. Their confidence intervals are similar but do not overlap with the interval of the answer ``none.'' Explanation E~9 relates to these proposition in context with the next question on impact analysis. As many respondents do impact analysis, explicit traces may make this activity more effective and efficient. 

\begin{figure}[!htb]
\centering\includegraphics[width=\textwidth]{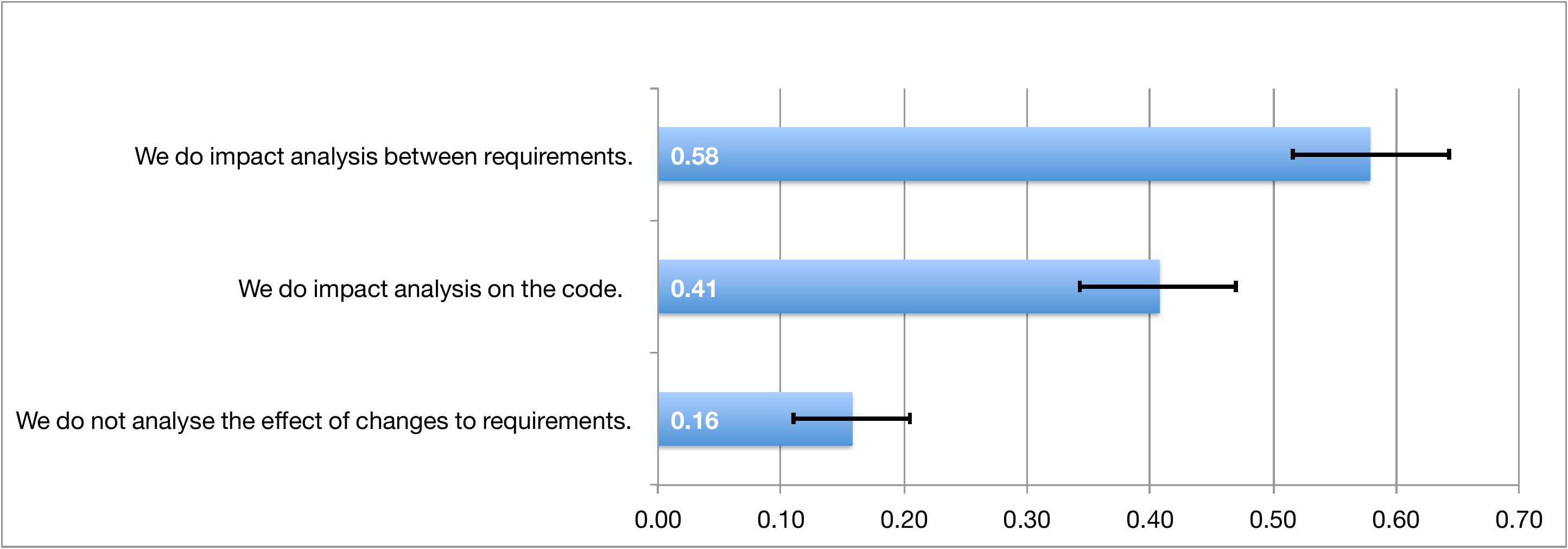}
\caption{How do you analyse the effect of changes to requirements? ($N = 228$)}\label{fig:changes}
\end{figure}

Figure~\ref{fig:changes} shows that impact analysis between requirements is done by the majority of respondents ($P = 0.58$ [0.51, 0.64]). Impact analysis on the code is still done by many respondents with $P = 0.41$ [0.35, 0.47]. No analysis of the effect of requirements changes is done only with $P = 0.16$ [0.11, 0.21].

Propositions P~19 and P~20 state that both impact analysis on the code and between requirements is done. Our data supports both propositions. They are the most often given answer and their confidence intervals do not overlap with the ``no analysis'' answer. The explanation E~10 discusses that despite that we saw in the question before many traces to the code, the impact of changes can still be seen best at the code level. Hence, we need both kinds of impact analyses. Yet, we notice that the two answers also have none-overlapping intervals. Hence, there are significantly more analyses between requirements. 


In addition, we added propositions on the status quo in aligning tests with requirements, which is critical to guarantee quality when requirements change and which gained increasing interest in the last years~\cite{bjarnason2014challenges}. These propositions are shown in Table~\ref{tab:p-alignment} and address the alignment measures of tester participation in requirements reviews, check of the coverage of requirements with tests, definition of acceptance criteria for requirements, as well as test derivation from system models.

\begin{table}[!htb]
\centering \scriptsize
\caption{Propositions about the status quo in aligning tests with requirements before the survey\label{tab:p-alignment}}{%
\begin{tabular}{lp{0.6\linewidth}ll}
\toprule
 & & \textbf{Supported in} & \textbf{Survey}\\
\textbf{No.} & \textbf{Propositions} & \textbf{first run or new} & \textbf{question}\\\midrule
P~21 & To align tests with requirements, testers participate in requirements reviews. & New & Q~15\\
P~22 & To align tests with requirements, the coverage of requirements with tests is checked. & New & Q~15 \\
P~23 & To align tests with requirements, acceptance criteria are defined for requirements. & New & Q~15 \\
P~24 & To align tests with requirements, tests are derived from system models. & New & Q~15 \\
\bottomrule
\end{tabular}}
\end{table}

\begin{figure}[!htb]
	\centering\includegraphics[width=\textwidth]{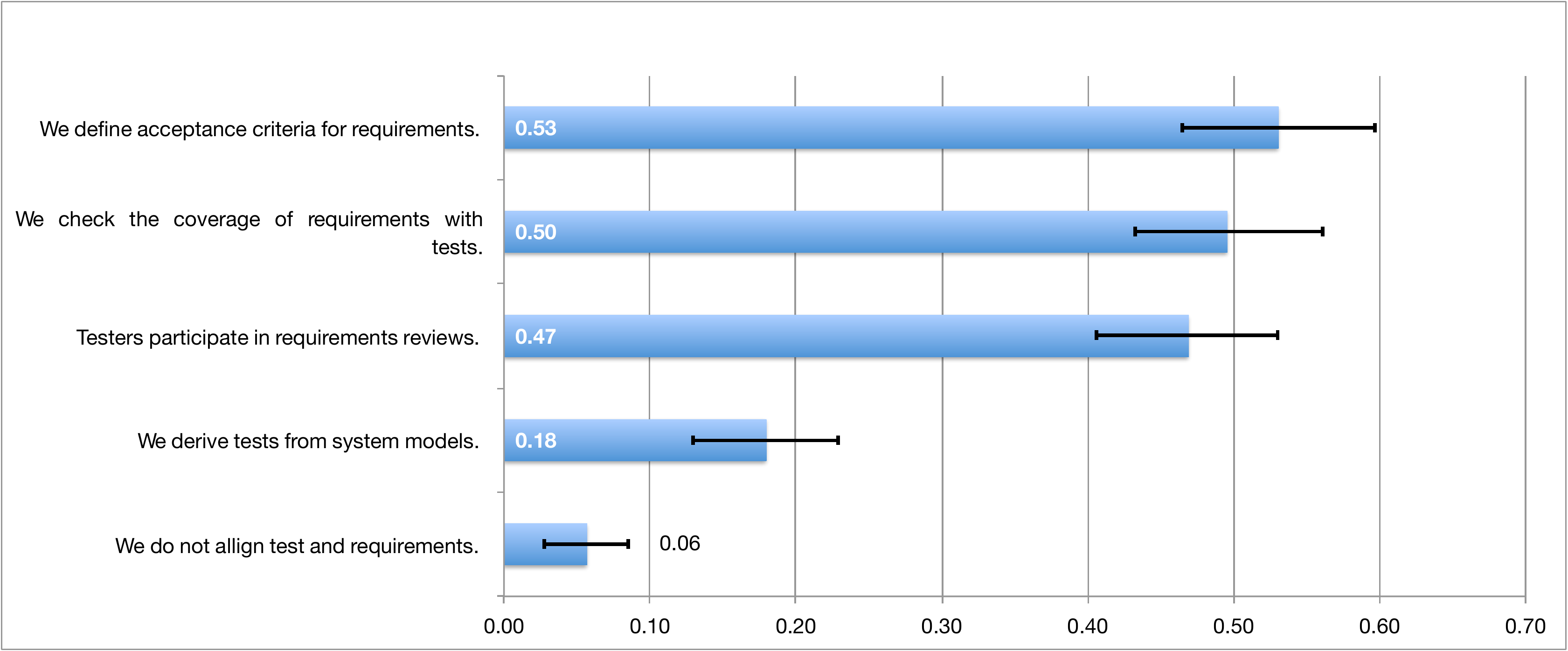}
	\caption{How do you align the software test with the requirements? ($N = 228$)}\label{fig:alignment}
\end{figure}

Figure~\ref{fig:alignment} shows that the three most frequent answers were that this is done via defining acceptance criteria ($P = 0.53$ [0.46, 0.60]), checking requirements coverage of tests ($P = 0.50$ [0.43, 0.56]) and testers participating in requirements reviews ($P = 0.47$ [0.41, 0.53]). Much more seldom are tests derived from system models ($P = 0.18$ [0.13, 0.23]). Only few of the respondents do not align tests and requirements ($P = 0.06$ [0.03, 0.09]).

The corresponding propositions P~21 to P~24 state that all the four possibilities besides not aligning requirements and tests are common in practice. Proposition P~24 on system models cannot be supported by our data. Only 18~\% of the respondents derive tests from system models. We will replace this proposition with: ``Deriving tests from system models is not used to align requirements and tests.'' Propositions P~21, P~22 and P~23 are supported in the data. Their confidence intervals strongly overlap and they are all used by about half of the respondents.

Table~\ref{tab:e-alignment} explains these propositions by the circumstance that several organizational and artefact-based measures are necessary to fully align tests with requirements. The lack of support for the old P~24 could be explained by the lack of existing system models that are complete or formal enough to derive tests (E-12).

\begin{table}[!htb]
\centering \scriptsize
\caption{Explanations for the propositions about aligning tests with requirements\label{tab:e-alignment}}{%
\begin{tabular}{lp{0.7\linewidth}l}
\toprule
\textbf{No.} & \textbf{Propositions} & \textbf{Changed} \\\midrule
P~21 & To align tests with requirements, testers participate in requirements reviews. &  \\
P~22 & To align tests with requirements, the coverage of requirements with tests is checked. &  \\
P~23 & To align tests with requirements, acceptance criteria are defined for requirements. &  \\
P~24 & Deriving tests from system models is not used to align requirements and tests. & \checkmark \\ \midrule
\textbf{No.} & \textbf{Explanations} & \textbf{Propositions} \\\midrule
E~11 & To fully align tests with requirements, organizational and artifact-based measures are necessary to link requirements and tests. & P 21--P 23\\
E~12 & Often, there are no system models that are complete or formal enough to derive tests. & P~24\\
\bottomrule
\end{tabular}}
\end{table}

Requirements changes are most critical if not propagated and traced accordingly. Defining test scenarios and test cases based on requirements (and updates according to requirement changes) can help to (a) better understand requirements and (b) check the expected behavior of the system in later testing phases. These test cases can be seen as ``definition of acceptance criteria'', i.e., if test cases are successfully executed and the requirement and/or requirement change has been understood and implemented correctly. Furthermore, test cases can be used to check whether or not requirements and/or requirement changes are in line with customer expectations. Thus, we can support the results of \citeN{CNV09} to propose test cases as criteria for acceptance tests. The activity ``define acceptance criteria'' can be seen as part of a change management process to elicit most valuable and correct requirements -- a common answer in our survey. 
In model-driven software engineering ~\cite{brambilla2017}, models often represent the foundation for software design, less frequently for code generation or test case generation. 

In practice, models seem not to be complete or formal enough to derive test cases. However, in research there is a strong focus on model-based testing and formal approaches when it comes to the alignment of requirements specification and testing as a recent systematic mapping study shows~\cite{barmi2011alignment}. However, limitations in practice often include the high effort for creating and maintaining models as foundation for deriving code and test cases. Thus, there is a trade-off between required efforts for model handling and benefits regarding frequent changes that need to be considered on business level. The straight-forward application of test cases in context of requirements and requirement changes can be seen as a first step in context of RE improvement.

\subsection{Status Quo in Requirements Engineering Process Standards}

We generally expect a company to have established a standardised way of working regardless of whether it is explicitly captured in a specific reference model or not. When launching NaPiRE, we worked based on the assumption that companies have established an explicit standard, mainly because we launched NaPiRE in Germany where we observed a strong standardisation in industry (e.g.\ in the automotive sector).

Also based on our observations, however, many companies have established their own standards and we generally expect a company-specific RE standard to be immature compared to standards for other disciplines due to the inherently complex nature of RE. We rely, for example, on the observations of Hall, Beecham and Rainer~\cite{HBR02} and suppose the standards to define coarse processes rather than, for instance, well defined artefact models that support traceability~\cite{RJ01}. 

In consequence, we expect the application of their standards to be neither practiced nor to be mandatory. 
This was actually one observation from our first NaPiRE run. 
Both form our new propositions as indicated in Tab.~\ref{tab:p-application}. 

\begin{table}[!htb]
\centering \scriptsize
\caption{Propositions about the status quo in the application and tailoring of requirements
engineering process standards before the survey\label{tab:p-application}}{%
\begin{tabular}{lp{0.6\linewidth}ll}
\toprule
 & & \textbf{Supported in} & \textbf{Survey}\\
\textbf{No.} & \textbf{Propositions} & \textbf{first run or new} & \textbf{question}\\\midrule
P~25 & Requirements engineers use their own RE standard. & New & Q~16 \\
P~26 & The RE standard is neither mandatory nor practiced. & New & Q~19\\
P~27 & The application of the RE standard is controlled via analytical quality assurance. & Supported & Q~20\\
P~28 & The RE standard is tailored at the beginning of a project by the project lead based on experiences. & Supported & Q~22\\
\bottomrule
\end{tabular}}
\end{table}

\begin{figure}[!htb]
	\includegraphics[width=\textwidth]{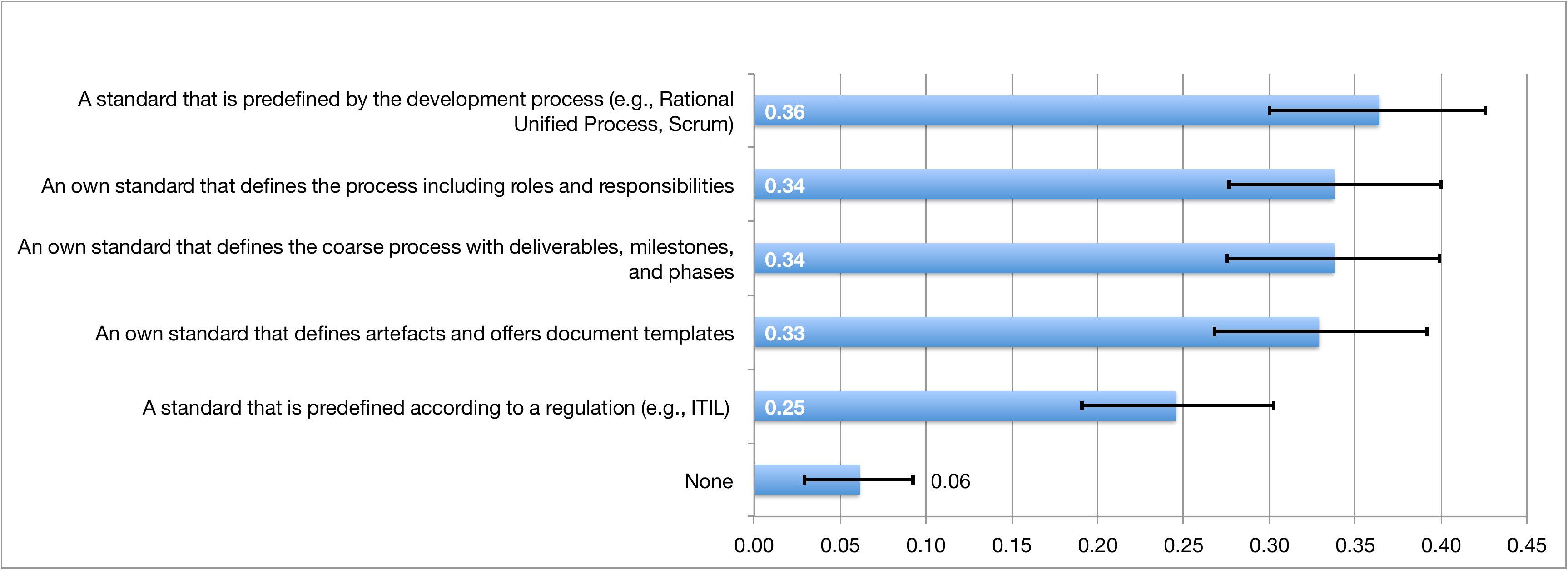}
	\caption{What RE company standard have you established at your company? ($N = 228$)}\label{fig:comp-standards}
\end{figure}

In the survey, we asked our respondents what RE company standard is established in their context. The proportions of answers with their CI is shown in Fig.~\ref{fig:comp-standards}. Most respondents stated that they use a standard predefined by the development process ($P = 0.3$6 [0.30, 0.43]). Also an internal standard that defines the process including roles and responsibilities ($P = 0.34$ [0.28, 0.40]), an internal standard that defines deliverables, milestones and phases ($P = 0.34$ [0.28, 0.40]), as well as an internal standard that defines artefacts and offers document templates ($P = 0.33$ [0.27, 0.39]) are common in our sample. Predefined standards according to a regulation like ITIL are less common in our sample ($P = 0.25$ [0.19, 0.30]). Using no standard at all or other standards tends not to be the case ($P = 0.06$ [0.03, 0.09]).

Proposition P~25 stating that requirements engineers use their own RE standard is supported by the data which indicates that an internal standard that defines the process including roles and responsibilities, an internal standard that defines deliverables, milestones and phases, as well as an internal standard that defines artefacts and offers document templates are common. They all have confidence intervals that are well beyond the threshold and do not overlap with the confidence interval of ``None.'' Yet, we notice that all of them overlap with the answer ``A standard that is predefined according to a regulation.'' Hence, this is reason for doubt whether an internal standard is more common in the population of software development organisations. Yet, as the answer about regulation goes below the 20~\% threshold, we do not define a new proposition for it.

Proposition P~25 can be explained by that RE practices generally need adaptation to fit the particularities of the context \cite{wohlrab18} (E~13). This explanation, however, is certainly still true for a broad range of phenomena, ranging from the adaptation of general RE strategies in dependency to the overall project setting to the choice of modelling techniques in dependency to the domain and the organizational culture of the development teams. Hence, it generally supports what has been commonly accepted in the RE research community~\cite{nuseibeh2000requirements}. A more detailed investigation that distils which context factors exactly affect which phenomena in which way is therefore necessary and in scope of future replications.

\begin{table}[!htb]
\centering \scriptsize
\caption{Propositions about the application and tailoring of requirements
engineering process standards with explanations after the survey \label{tab:e-application}}{%
\begin{tabular}{lp{0.7\linewidth}l}
\toprule
No. & \textbf{Propositions} & \textbf{Changed} \\\midrule
P~25 & Requirements engineers use their own RE standard. &  \\
P~25a & Requirements engineers use a standard that is predefined by the development process. & \checkmark\\
P~26 & RE standards are practised regardless whether they are mandatory or not. & \checkmark \\
P~27 & The application of the RE standard is controlled via analytical quality assurance. &  \\
P~27a & The application of requirements engineering standards is checked by project assessments & \checkmark \\
P~27b & The application of requirements engineering standards is checked by constructive quality assurance (e.g.\ via checklists or templates). & \checkmark\\
P~28 & The RE standard is tailored at the beginning of a project by the project lead based on experiences. &  \\ \midrule
\textbf{No.} & \textbf{Explanations} & \textbf{Propositions} \\\midrule
E~13 & Requirements engineering differs quite strongly over domains and contexts and, thus, needs to adapt to these to be effective. & P~25, P~26\\
E~14 & Many practiced development processes prescribe or are associated with a specific way of performing requirements engineering. & P~25a\\
E~15 & The project lead knows the specific of the domain and project context best. & P~28\\
\bottomrule
\end{tabular}}
\end{table}

The most frequent answer, however, was that a standard predefined by the development process is established. We add this as new proposition P~25a to our theory. An explanation could be that many respondents use development processes such as Scrum or V-Model XT that contain or are associated with a specific way of dealing with requirements engineering (E~14). For example, Scrum or other agile projects are usually expected to have some kind of product owner, a backlog and user stories.

\begin{figure}[!htb]
	\includegraphics[width=\textwidth]{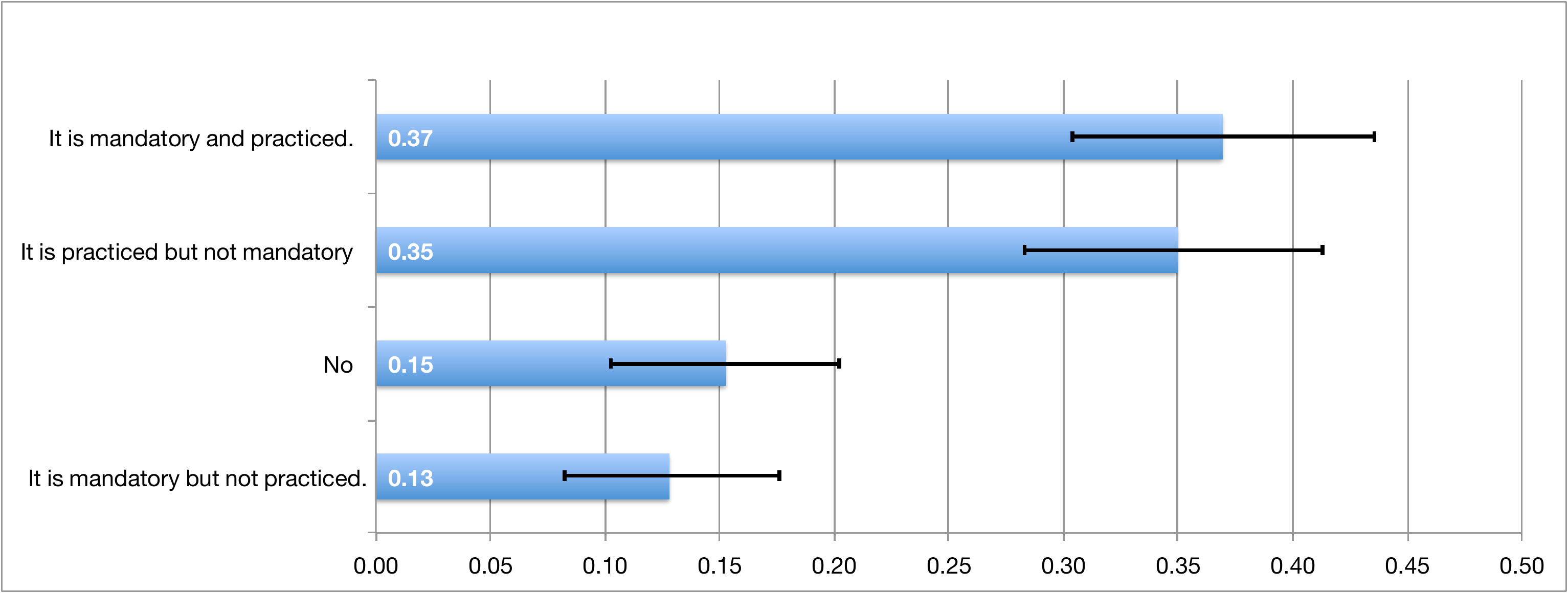}
	\caption{Is the requirements engineering standard mandatory and practised? ($N = 203$)}\label{fig:standard-mandatory}
\end{figure}

With regards to whether a requirements engineering standard is mandatory or practised (see Fig.~\ref{fig:standard-mandatory}), most respondents answer that the standard is mandatory and practised ($P = 0.37$ [0.30, 0.43]) or practised but not mandatory ($P = 0.35$ [0.29, 0.42]). 

Proposition P~26 stating that requirements engineering standards are neither practised nor mandatory is not supported by the data. Quite to the contrary, it seems that the standards are practised in most organisations regardless whether they are mandatory or not. The two answers that the standard is practised have overlapping confidence intervals that do not cross the other two answer possibilities. Only $P = 0.15$ [0.10, 0.20] stated that their standard is neither mandatory nor practised. Even less ($P = 0.13$ [0.08, 0.17]) answered that the standard is mandatory but not practised. Hence, we  replace P~26 by ``Company standards are practised regardless whether they are mandatory or not.''

\begin{figure}[!htb]
\includegraphics[width=\textwidth]{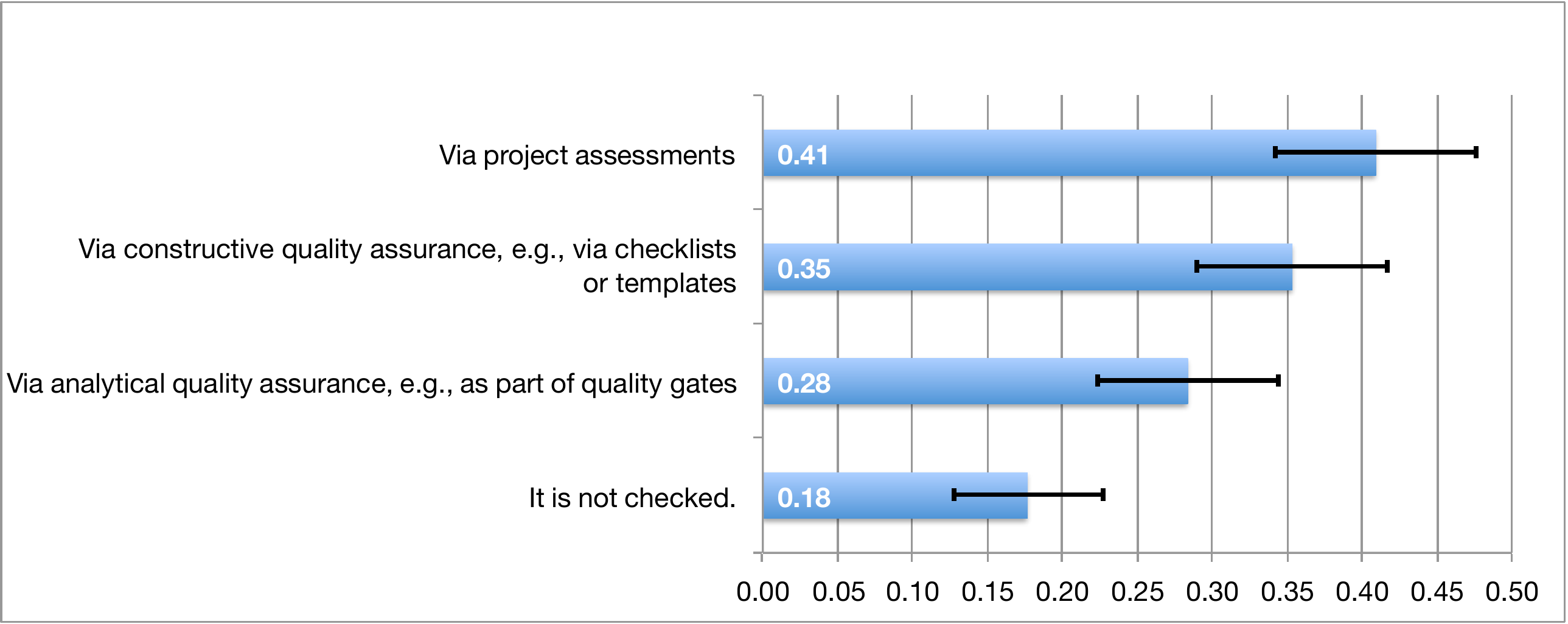}
\caption{How do you check the application of your requirements engineering standard? ($N = 215$)}\label{fig:standard-application-check}
\end{figure}

Furthermore, we asked the respondents how the application of requirements engineering standards is checked (see Fig.~\ref{fig:standard-application-check}). The most common way to do so is via project assessment with P = 0.41 [0.23, 0.48]. Moreover, constructive quality assurance ($P = 0.35$ [0.29, 0.42]) but also analytical quality assurance ($P = 0.28$ [0.22, 0.34]) are common in our sample. Not checking requirements also occurs bot not commonly ($P = 0.18$ [0.13, 0.23]).

Proposition P~27 stating that the application of the RE standard is controlled via analytical quality assurance can be supported by the data. Yet, project assessments and constructive quality assurance are applied more frequently in our sample than analytical quality assurance. With 28~\% of the respondents stating that they use analytical quality assurance, we will keep the proposition in the theory but add additional propositions for the other two methods: ``The application of requirements engineering standards is checked by project assessments.'' and ``The application of requirements engineering standards is checked by constructive quality assurance (e.g.\ via checklists or templates).''

\begin{figure}[!htb]
\includegraphics[width=\textwidth]{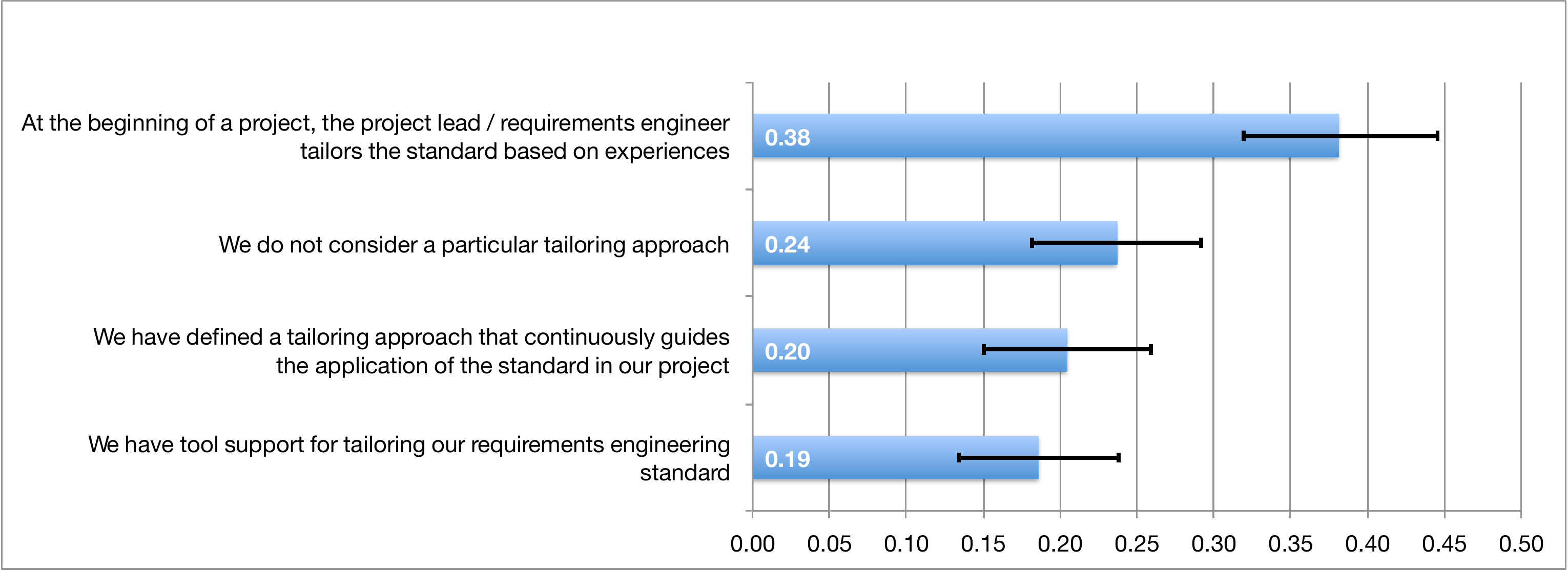}
\caption{How is your requirements engineering standard applied in your regular projects? ($N = 215$)}\label{fig:standard-application}
\end{figure}

With regards to how requirements engineering standards are applied in regular projects (see Fig.~\ref{fig:standard-application}), tailoring based on experience is most common in our sample with $P = 0.38$ [0.32, 0.44]. Not considering a particular tailoring approach ($P = 0.24$ [0.18, 0.29]), having a tailoring approach that continuously guides the application of the standard ($P = 0.20$ [0.14, 0.26]) as well as having tool support for tailoring requirements engineering standards ($P = 0.19$ [0.13, 0.24]) are moderately applied. 

Proposition P~28 on RE standard tailoring by the project lead based on experience is clearly supported by the data. It is the most common answer and has a confidence interval that does not overlap with the other remaining answers. This could be explained by the project lead having the best knowledge of the domain and context of the project (E~15).


Table~\ref{tab:p-reasons} summarises all our propositions on reasons and barriers for defining a requirements engineering process standard. 
We expect that there is a high diversity in reasons why companies establish a process standard.  Our first NaPiRE run confirmed nine propositions which we kept. We reversed six that were not supported and kept one although it was not supported.

\begin{table}[!htb]
\centering \scriptsize
\caption{Propositions about the reasons and barriers for defining a requirements engineering process standard before the survey\label{tab:p-reasons}}{%
\begin{tabular}{lp{0.6\linewidth}ll}
\toprule
 & & \textbf{Supported in} & \textbf{Survey}\\
\textbf{No.} & \textbf{Propositions} & \textbf{first run or new} & \textbf{question}\\\midrule
P~29 & Compliance to regulations and standards (like CMMI) does not motivate a standard. &  Opposite not supported & Q~17\\
P~30 & Seamless development by integrating RE into the development process motivates a standard. & Supported & Q~17 \\
P~31 & Better tool support motivates a standard. & Supported & Q~17\\
P~32 & Formal prerequisites for project acquisition do not motivate a standard. & Supported & Q~17\\
P~33 & Support of distributed development motivates a standard. & Not supported & Q~17 \\
P~34 & Support of progress control motivates a standard. & Opposite not supported & Q~17 \\
P~35 & Better quality assurance of artefacts motivates a standard. & Supported & Q~17\\
P~36 & Support of benchmarks does not motivate a standard. & Opposite not supported & Q~17\\
P~37 & Support of project management and planning motivates a standard. & Supported & Q~17 \\
P~38 & Higher efficiency motivates a standard. & Supported & Q~17\\
P~39 & Knowledge transfer motivates a standard. &  Supported & Q~17\\
P~40 & Higher process complexity barriers defining a standard.  & Supported & Q~18\\
P~41 & Higher demand for communication barriers defining a standard. & Opposite not supported & Q~18\\
P~42 & Lower efficiency does not barrier defining a standard. & Opposite not supported & Q~18\\
P~43 & Missing willingness to change barriers defining a standard. & Supported & Q~18\\
P~44 & Missing possibilities of standardisation does not barrier defining a standard. & Opposite not supported & Q~18\\ \bottomrule
\end{tabular}}
\end{table}

\begin{figure}[!htb]
	\includegraphics[width=\textwidth]{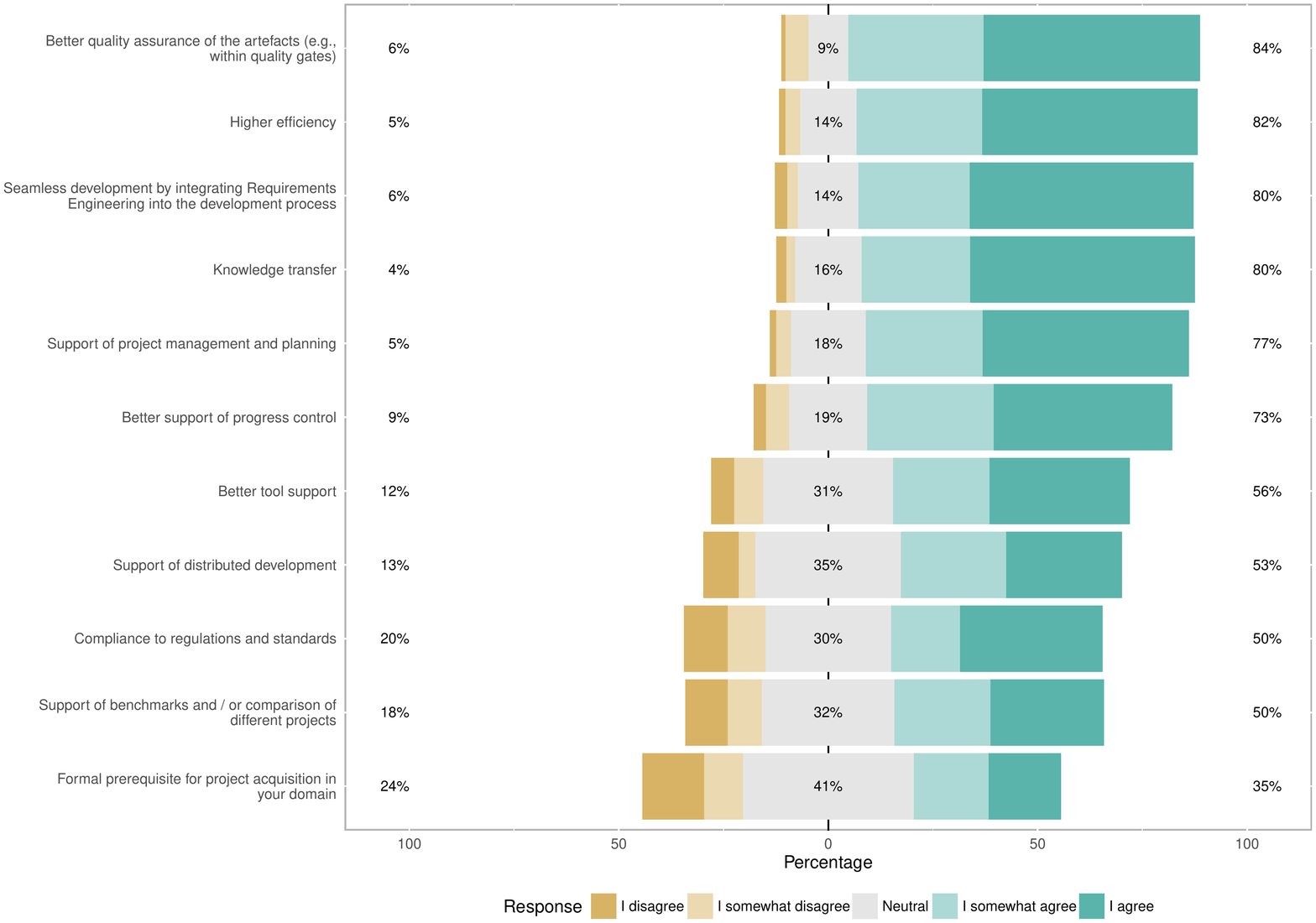}
	\caption{Which reasons do you agree with as a motivation to define a company standard for RE in your company? ($N = 228$)}\label{fig:motivation-standards}
\end{figure}

In the following $M$ denotes the mean of the corresponding data.
Main motivations to define a company standard for RE (see Fig.~\ref{fig:motivation-standards}) are better quality assurance of artefacts (median = 4, $M = 4.3$ [4.2, 4.4]), higher efficiency (median = 4, $M = 4.3$ [4.1, 4.4]), seamless development by integration of RE into the development process (median = 4, $M = 4.3$ [4.1, 4.4]), knowledge transfer (median = 5, $M = 4.3$ [4.1, 4.4]), support of project management and planning (median = 4, $M = 4.2$ [4.1, 4.3]) and support of progress control (median = 4, $M = 4.0$ [3.9, 4.2]). Still more agreement than disagreement have better tool support (median = 4, $M = 3.7$ [3.5, 3.9]), support of distributed development (median = 3, $M = 3.6$ [3.4, 3.8]), compliance to regulations and standards (median = 3, $M = 3.5$ [3.3, 3.7]), support of benchmarks and/or comparison of different projects (median = 3, $M = 3.5$ [3.3, 3.7]) and formal prerequisite for project acquisition in the participant's domain (median = 3, $M = 3.1$ [3.0, 3.3]). 

The results support our theory only partially. We have clear support with a median and lower boundary of the mean CI above 3 for the positive propositions P~30, P~31, P~34, P~35 and P~37--P~39. We do not have clear support for P~33 and the negative propositions P~29, P~32 and P~36 as their medians are all 3 and the confidence intervals of the means are all at 3.0 or above. Hence, we remove them from the theory for now. It might be possible to reintroduce them with more specific context. For example, there is some indication in the data that in organisations working for the public sector, in the automotive, avionics or finance domain, compliance to regulations and standards are more important motivators than for other organisations.

\begin{table}[!htb]
\centering \scriptsize
\caption{Propositions about the reasons and barriers for defining a requirements engineering process standards and explanations after the survey \label{tab:e-reasons}}{%
\begin{tabular}{lp{0.8\linewidth}l}
\toprule
\textbf{No.} & \textbf{Propositions } & \textbf{Changed} \\\midrule
\sout{P~29} & \sout{Compliance to regulations and standards (like CMMI) does not motivate a standard.}    & \checkmark  \\
P~30 & Seamless development by integrating RE into the development process motivates a standard. &  \\
P~31 & Better tool support motivates a standard. & \\
\sout{P~32} & \sout{Formal prerequisites for project acquisition do not motivate a standard.} & \checkmark\\
P~33 & Support of distributed development motivates a standard. &  \\
P~34 & Support of progress control motivates a standard. &  \\
P~35 & Better quality assurance of artefacts motivates a standard. & \\
\sout{P~36} & \sout{Support of benchmarks does not motivate a standard.} & \checkmark \\
P~37 & Support of project management and planning motivates a standard. &  \\
P~38 & Higher efficiency motivates a standard. & \\
P~39 & Knowledge transfer motivates a standard. &  \\
P~40 & Higher process complexity barriers defining a standard.  & \\
\sout{P~41} & \sout{Higher demand for communication barriers defining a standard.} & \checkmark \\
P~42 & Lower efficiency does not barrier defining a standard. & \\
P~43 & Missing willingness to change barriers defining a standard. & \\
\sout{P~44} & \sout{Missing possibilities of standardisation does not barrier defining a standard.} & \checkmark \\ 
\midrule
\textbf{No.} & \textbf{Explanations} & \textbf{Propositions}\\\midrule
E~16 & An RE standard can help to integrate RE activities and artefacts with other development activities and artefacts. & P~30\\
E~17 & It is more efficient to build or acquire tool support for RE if the activities and artefacts are standardised. & P~31\\
E~18 & Standardised RE artefacts make it easier to check if they are created and, hence, support progress control. & P~34\\
E~19 & If RE artefacts are standardised, standardised QA can be used such as checklists or automatic checks. & P~35\\
E~20 & If the project lead can rely on a standardised RE, the planning can rely on the standardised activities and artefacts. & P~37\\
E~21 & A standardised RE allows the project participants to become experts in it and, therefore, become more efficient. & P~38\\
E~22 & The RE standard codifies good practices and experiences which can be transferred to new projects and project
	participants. & P~39\\
E~23 & An RE standard might force projects to a more complex RE process than necessary for the concrete context. & P~40\\
E~24 & Using RE process standards is considered more efficient (see also P~38). & P~42\\
E~25 & People in general are resistant to change. & P~43\\
\bottomrule
\end{tabular}}
\end{table}

\begin{figure}[!htb]
	\includegraphics[width=\textwidth]{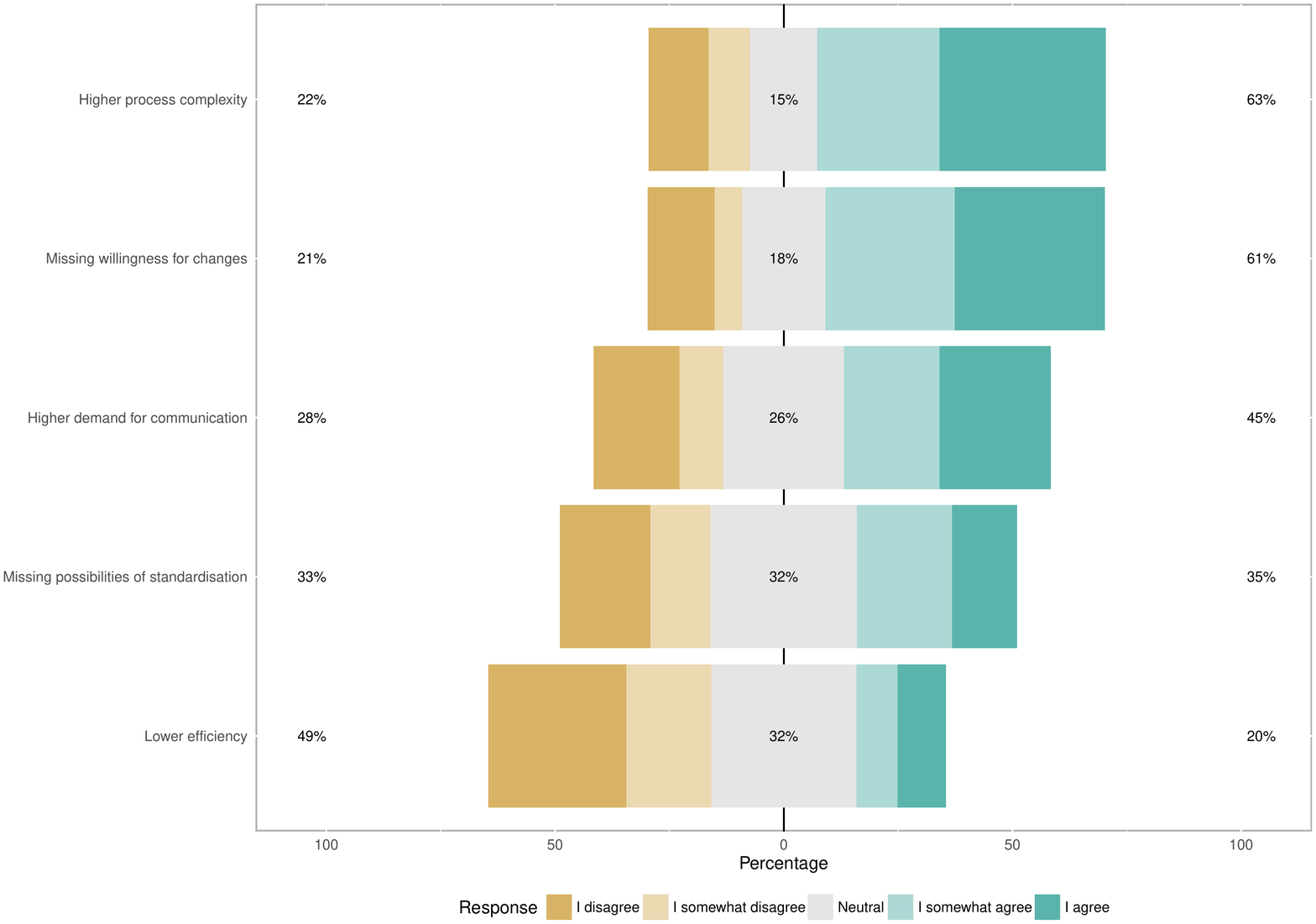}
	\caption{Which reasons do you agree with as a barrier to define a company standard for RE in your company? ($N = 228$}\label{fig:barrier-standards}
\end{figure}

Main barriers to define a company standard for RE (see Fig.~\ref{fig:barrier-standards}) are higher process complexity (median = 4, $M = 3.6$ [3.4, 3.8]) and missing willingness for changes (median = 4, $M = 3.6$ [3.4, 3.8]). Higher demand for communication (median = 3, $M = 3.2$ [3.0, 3.4) and  missing possibilities for standardisation (median = 3, $M = 3.0$ [2.8, 3.1]) are more mixed. For lower efficiency (median = 2, $M = 2.5$ [2.3, 2.7]), we even have more respondents disagreeing than agreeing. 

Propositions P~40 to P~44 state whether specific barriers exist in practice or not. Our results support all propositions apart from P~41 and P~44 that higher demand for communication or missing possibilities of standardisation barriers defining a standard. Both are on average close to the neutral answer. Similarly as above, we will remove them for now from the theory and will look for further context factors where they might have stronger agreement. 

In Table~\ref{tab:e-reasons} we provide explanations affecting all supported propositions on motivations and barriers on defining RE standards. We derived several of the explanations from answers to various open questions in the questionnaire and added existing work where appropriate. We may expect that standardisation of RE activities and artefacts would support a process integration as standards make explicit all relevant concepts and dependencies in the elements of a process (e.g., which information in use case models is required to support testing activities)~\cite{fernandez2015artefact} (E~16). It is reasonable to have similar positive expectations for planning, quality assurance, and progress control of the artefacts, because a certain (context-specific) standardisation of the RE artefacts allows for the definition of quality criteria or certain degrees of completion of artefacts (E~18, E~19). Standardisation of RE, however, is generally considered difficult, because practices considered useful in one context could be perceived completely alien to the culture and needs of the next. Hence, standardisation of RE might enforce practices that do not fit the context, thus, making the RE process more complex than necessary -- especially when relying on universal standardisation norms~\cite{pettersson2008practitioner} (E~23). This might be also one reason why one of the major challenges in software process improvement is coping with people's resistance to change~\cite{kauppinen2004implementing} (E~25).

There is little existing research on requirements engineering standards. We have some support for the finding of Nikula, Sajaniemi and K\"alvi\"ainen~\citeN{nikula2000sps} that the majority of organisations explicitly define their individual RE process. To our knowledge, there is also no systematic literature review on requirements engineering standards.

\subsection{Status Quo in Requirements Engineering Improvement}

Process improvement is important for any software engineering practice but for an activity as volatile and complex as RE, we expect this to be essential. We skipped a more general question of how requirements engineering improvement is done because we already had support from the previous
run that improvement is done continuously. Our propositions are shown in Tab.~\ref{tab:hypothesisRQ3}.
The new propositions resulted from our observation in the first NaPiRE run that RE process improvement is conducted by dedicated roles and based on own established principles and approaches.

\begin{table}[!htb]
\centering \scriptsize
\caption{Propositions and explanations about requirements engineering improvement prior to the survey \label{tab:hypothesisRQ3}}{%
\begin{tabular}{lp{0.6\linewidth}ll}
\toprule
 & & \textbf{Supported in} & \textbf{Survey}\\
\textbf{No.} & \textbf{Propositions} & \textbf{first run or new} & \textbf{question}\\\midrule
P~45 & Requirements engineering is continuously improved. & Opposite not supported & Q~23\\
P~46 & A continuous improvement is done to determine strengths and weaknesses. & Supported & Q~24\\
P~47 & Requirements engineering is improved via an own business unit / role. & New & Q~23\\
P~48 & RE is improved by an internally defined standard. & New & Q~23\\ \bottomrule
\end{tabular}}
\end{table}

\begin{figure}[htb]
\includegraphics[width=\textwidth]{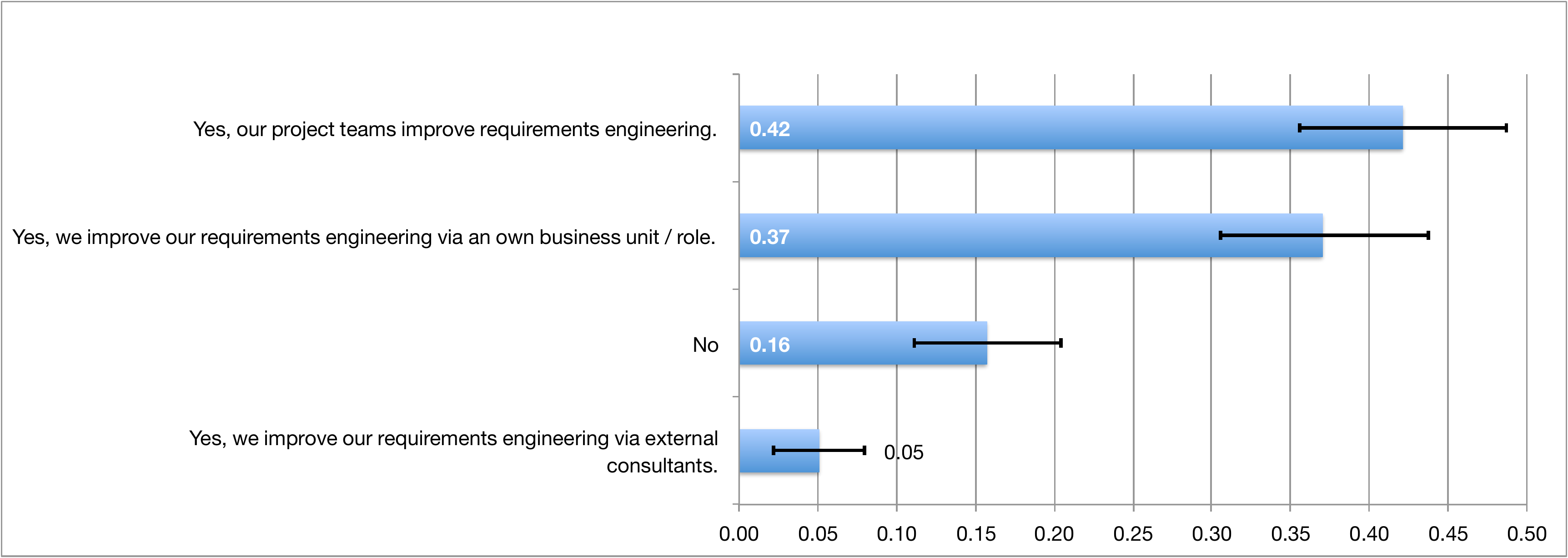}
\caption{Is your requirements engineering continuously improved? ($N = 216$)}\label{fig:is-continuous}
\end{figure}

In the survey, we asked whether the organisations improve their RE continuously and who is responsible for this improvement. The results in Fig.~\ref{fig:is-continuous} show project teams commonly improve requirements engineering with $P = 0.42$ [0.36, 0.49]. Requirements engineering is to a large extent also improved via an own business unit or role ($P = 0.37$ [0.30, 0.44]). No improvement is less common ($P = 0.16$ [0.11, 0.20]), and improvement via external consultants is rare ($P = 0.05$ [0.02, 0.08]). Proposition P~45, which states that requirements engineering is continuously improved, is supported by the data. Furthermore, proposition P~47 on RE improvement via an own business unit or role is also supported by data from Q~23 shown in Fig.~\ref{fig:is-continuous}. 

\begin{table}[!htb]
\centering \scriptsize
\caption{Propositions about requirements engineering improvement and explanations after the survey \label{tab:e-improvement}}{%
\begin{tabular}{lp{0.7\linewidth}cc}
\toprule
\textbf{No.} & \textbf{Propositions} & \textbf{Changed} \\ \midrule
P~45 & Requirements engineering is continuously improved. & \\
P~46 & A continuous improvement is done to determine strengths and weaknesses. & \\
P~47 & Requirements engineering is improved via an own business unit / role. &  \\
P~48 & RE is improved by an internally defined standard. &  \\ 
P~49 & RE is improved using external normative standards. & \checkmark \\\midrule
\textbf{No.} & \textbf{Explanations} & \textbf{Propositions} \\\midrule
E~26 & Many companies have realised the importance of requirements engineering and of
	continuous improvement of development processes and methods. Working on it continuously
	helps to not forget strengths and weaknesses of the current RE approach. & P45, P~46\\
E~27 & RE improvement is performed by internally defined standards and best supported by an own business unit or role. & P~47\\
E~28 & External normative standards are often considered too complex and elaborate to apply. & P~48\\
\bottomrule
\end{tabular}}
\end{table}

There are no studies investigating RE improvement directly comparable at this level of granularity. Yet, a systematic mapping study~\cite{pekar2014improvement} adds that quality assessment plays an important role for improving software requirements specifications.

\begin{figure}[htb]
\includegraphics[width=\textwidth]{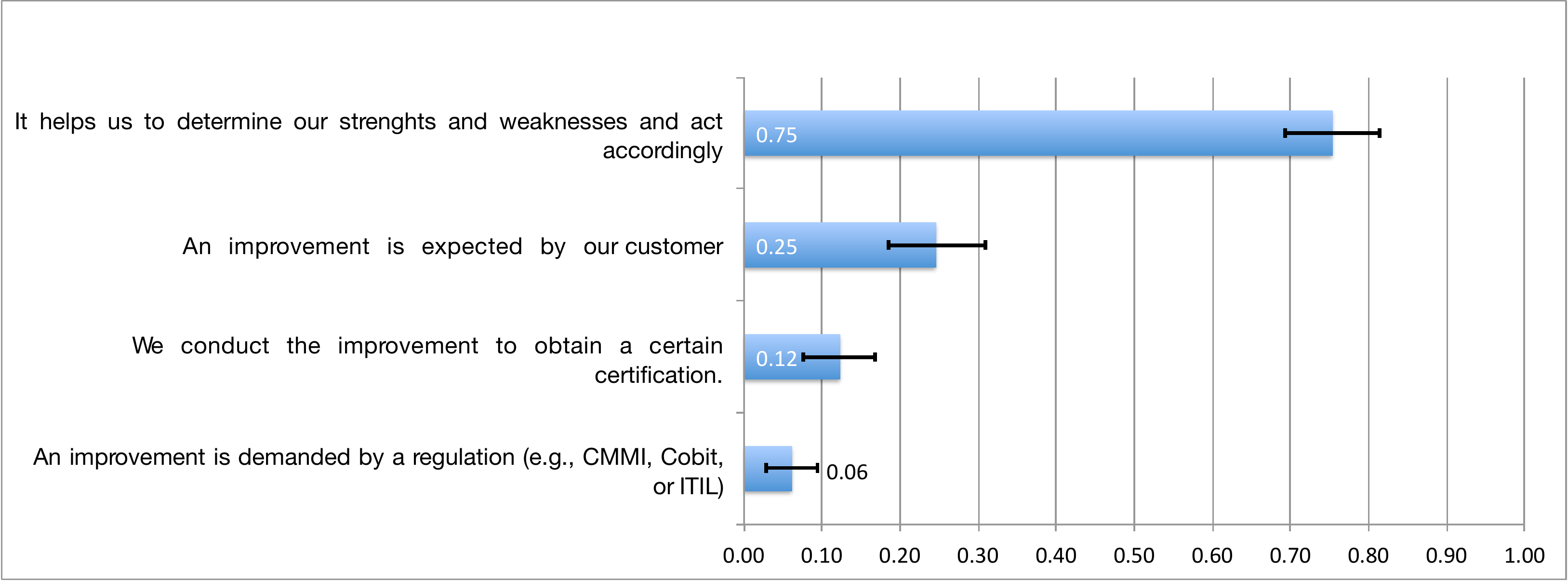}
\caption{Why do you continuously improve your requirements engineering? ($N = 195$)}\label{fig:why-continuous}
\end{figure}

At this point, we wanted to dig deeper and understand the reasoning behind doing a continuous improvement. As shown in Fig.~\ref{fig:why-continuous}, the most common and often applied improvement measure is to determine strengths and weaknesses with $P = 0.75$ [0.69, 0.81]. Sometimes improvement is driven by customers ($P = 0.25$ [0.18, 0.31]). Rarely, improvements are conducted to obtain a certain certification ($P = 0.12$ [0.08, 0.17]) or due to a regulation like CMMI, Cobit or ITIL ($P = 0.06$ [0.03, 0.10]). Additional other reasons include better efficiency, improvement of quality in project development and adoption of an agile method that has inspection as one of the principles and using them to promote continuous improvement.
Proposition P~46 stating that continuous improvement is done to determine strengths and weaknesses is supported by the data. E~25 explains both propositions P~45 and P~46.  

Finally, we asked whether the participants use a normative external standard such as CMMI for RE for the improvement. We found that most use an internally defined (company-specific) standard for improvement ($P= 0.63$ [0.56, 0.70]) but there are also many applying external standards ($P = 0.37$ [0.29, 0.44]) although, as we saw above, certification is not a major reason for improvements. We have support for P~48 that RE is improved by an internally defined standard. Yet, the data suggests that we should add a new proposition that RE is also improved using external normative standards. An explanation for P~48 is suggested by many textual answers to the follow-up question why an external standard is not used. We got answers such as ``There is no need for using an external standard. Therefore, we decided to use a lightweight internal standard.'', ``Simplicity :-)'' or ``we use very lean improvement process; less effort''. We interpret this such that external normative standards are often considered too complex and elaborate to apply. We document this as explanation E~28. 

Staples \textit{et al}.~\citeN{SNJABR07} had already found that smaller companies show a reluctance against normative improvement approaches. We can confirm this and extend it to also include larger companies. The reluctance seems to be a more general phenomenon.

\section{Conclusion}
\label{sec:Conclusion}

In this section we first present a summary of the results and relate them to existing evidence. We then discuss the impact and implications of the presented work, present limitations of the performed study, and finally discuss directions of future work.

\subsection{Summary of Results}

Surprisingly for us, we found no strong differences among surveyed countries and regions. The detailed analysis is not in the main results as we wanted to focus then on answering our research questions based on the evaluation of our theory. Nevertheless, initially it was one of the aspects we wanted to investigate in more detail. Some comparisons are published in \cite{mendez:softw15}. Yet, with the sample we have now, we cannot support significant difference between countries and regions.

\subsubsection{Requirements Elicitation and Documentation (RQ 1)}
Most of the respondents state to document requirements textually in free form or with some constraints. Only for use cases and data models, semi-formal approaches are used. Semi-formal and formal goal models seem to be niche. Non-functional requirements are more often quantified than not (maybe because of performance-related non-functional requirements classes).

\subsubsection{Requirements Change and Alignment (RQ 2)}
Change management is either continuous or after formally accepting a requirements specification. Less then 20~\% change the specification itself regularly. Most respondents update their backlog or only work with change requests.
Most respondents do impact analysis between requirements, many do impact analysis in the code. For that, many respondents use traces between requirements and code and requirements and design documents.
To align tests with requirements, many respondents define acceptance criteria, cover requirements with tests and let testers participate in requirements reviews. Only 18~\% derive tests from system models.

\subsubsection{Requirements Engineering Standards (RQ 3)}
Almost all respondents have an RE company standard. There is a similar share of different kinds of standards in the organisations of our respondents. A standard is also practised by most of the organisations. In many organisations, the standard is tailored at the beginning of the project by the project lead based on experience. Several methods are in use to check the application of the standard.
There are various reasons to define an RE standard. Better quality assurance and higher efficiency are common motivations while it is least often a formal prerequisite for project acquisition. Main barriers for defining a standard are higher process complexity because of the standard, missing willingness for change and higher demand for communication.

\subsubsection{Requirements Engineering Improvement (RQ 4)}
Most organisations improve their requirements engineering continuously either by the project teams themselves or by an own business unit/role. The motivation for the improvement is overwhelmingly internally driven: It helps the organisations to determine their strengths and weaknesses and to act accordingly. Improvements for certifications and demanded by regulations are rarely the motivator.

\subsection{Comparison to Last Run}

In our theory from the first run \cite{MW14}, we covered some aspects that we did not include in this run of the survey. Hence, we cannot compare the new results with the old results on expectations on good requirements engineering and details of problems with their requirements engineering standards. Furthermore, we published the discussion of common problems and their causes and effects based on the second run separately~\cite{napire:emse15}.

Overall, we have 22 propositions that were already covered in the first run of the survey. 
They were presented together with their outcome from the first run in section~\ref{sec:Results} with the theory of their respective research questions. Either we ran the same proposition in the first run, or in case it was not supported there, we might have had the opposite proposition. For the majority of propositions, the second run confirms the results of the first run. We have 14 propositions that were supported in both runs directly. For three propositions, we had the opposite proposition not supported in the first run and could support them now in the second run. 

There are two further types of situations: (1) The proposition P~33, that the support of distributed development motivates a standard, was not corroborated by the data in the first run. Based on our own experience, we nevertheless kept the original proposition. Yet, the data of the second run now did not clearly support it either with an average close to neutral. (2) For P~29, P~36, P~41 and P~44, we had the opposite in the first run which was not supported by the data. We changed them for the second run but could also not support them. They received also no clear agreement or disagreement with averages close to neutral. We decided to remove these unclear propositions from the theory for now as is reflected in the revised theory after the survey in section~\ref{sec:Results}.

\subsection{Impact/Implications}
We hope the impact of our work is twofold: First, practitioners can use our results to compare their own experiences and practices to the status quo in practice more generally. This might help them for introspections of their processes and techniques, lead to improvement initiatives and maybe to try out new techniques.

Second, we have now a theory that (in parts) unified previously isolated ones and that has been validated by two successive survey runs. Hence, it constitutes the most solid foundation to how requirements engineering is done in practice so far. Therefore, we believe it can be the starting point for a variety of further research. Researchers working on specific techniques can check the usage of their techniques in practice and reflect on this usage. In particular, semi-formal and formal goal models are rarely used despite a considerable attention in research. Hence, it should be further investigated whether this is caused by deficiencies in these methods or just insufficient knowledge on the side of the practitioners. Many other aspects of the theory (and extensions of the theory) should be investigated with further surveys and, especially, with other research techniques such as interviews and case studies. This would help to better understand why the status quo is like it is now.

\subsection{Limitations}

\subsubsection{Conclusion Validity}
We were not able to construct a random sample systematically covering different types of organisations and requirements engineers, as we do not have general information about how the population as such looks like. To deal with this limitation, we used bootstrapping and only employed confidence intervals instead of null hypothesis testing to evaluate the propositions of our theory.

Moreover, the choice of the thresholds we used (20~\% and above ``neutral'') is to some degree arbitrary. They are based on our first run and discussions in the NaPiRE team. Yet, there could be reasonable cases for other numbers.

\subsubsection{Internal Validity}
A limitation that we always have with survey research is that surveys can only reveal stakeholders' perceptions on current practices in a cross-sectional manner. These perceptions might further be distorted and not fully represent reality. Also related to that, the different respondents might have interpreted different terms and concepts in the questionnaire differently in dependency to their organizational environment; for instance, the notion of "RE standard [process/approach]". We tried to minimize the resulting threats to validity via the joint community workshops described in section ~\ref{sec:NaPiREInitiative} and the subsequent industrial pilot studies, yet, we cannot completely eradicate the possibility of misinterpretations, misconceptions or simple bias. We believe that we need to live with this limitation for now to establish an empirical basis that afterwards can be complemented with other research methods to analyse those perceptions in more detail.

The analysis of the ``other'' options in the closed questions did not reveal any frequent answers that we missed in our answer options. It might still be the case that the respondents were not motivated to spend the time to fill in free-text answers. Yet, we believe that we could minimise this threat.

\subsubsection{Construct Validity}
Furthermore, requirements engineering is a broad field which we cannot realistically cover in a single survey. While we aimed to be rather broad in our questions, many details could not be completely covered, such as more detailed techniques for requirements elicitation or other stakeholders such as customers or users. This is one of the reasons why we set up NaPiRE to be regularly repeated. We use this to shift the focus on different aspects to refine our theory while we are able to retest other parts of the theory at the same time. A specific unfortunate omission for this run of the survey was that we did not include user stories explicitly (only as part of textual requirements lists with constraints) although we incorporated several other agile concepts such as the product backlog explicitly.

\subsubsection{External Validity}
Being only able to participate by invitation might introduce the threat that the participants are biased in the sense that they are somehow known to members of the NaPiRE team. Yet, in some countries, we used specific mailing lists for which we had clear knowledge that our invitation criteria are met. Furthermore, as discussed above in the design, we weight the benefits of control over the list of invitees as higher than this risk.

Finally, a bigger shortcoming is that we still cannot claim to cover requirements engineering as done in the whole world. While we have a large and geographically distributed sample with response rates between 25 and 75~\%, our respondents primarily come from Europe, North and South America. This probably covers a considerable part of the world's software companies. Nevertheless, to be able to really generalize and potentially see differences, we would need to cover Asia, Africa, and Australia as well. Especially Asia has now a strong software industry and potentially stronger cultural differences to our current sample. For example, the level of trust in organizations differ depending on whether it is in a individualist or collectivist society~\cite{huff2003levels}.

\subsection{Future Work}
We are committed to further runs of the survey in the NaPiRE initiative. In parallel to this analysis, we are finalising the third run of the survey which will significantly add details on the handling of different types of quality requirements or the relationship between the software teams and the customers. Future studies should include even more theory considering the customer and user. If possible, it would also be interesting to compare difference across application domains.

In addition, as we have so far concentrated on purely descriptive propositions, we now move to more causal propositions on the relationship between different factors, such as different elicitation practices and problems. Furthermore, we are planning studies that employ other research methods, as discussed above, to get other viewpoints on the perceptions that we capture with our surveys. Finally, we will also integrate our theory with other related theories like the theory of distances in software engineering~\cite{bjarnason2016theory} as both theories do not conflict and the used concepts overlap only partially.

\begin{acks}
The authors would like to thank all practitioners who took the time to respond to our survey as well as all colleagues who have supported the NaPiRE initiative along the way, including the International Requirements Engineering Board for their financial support during the analysis of the recent (third) run.
Tayana Conte is supported by CNPq (311494/2017-0).
Dietmar Pfahl was supported by the institutional research grant IUT20-55 of the Estonian Research
Council. Rafael Prikladnicki is partially funded by Fapergs (process 17/2551-0001205-4) and CNPq.
For the work of Dietmar Winkler, the financial support by the Austrian Federal Ministry of Science, Research and Economy and  the National  Foundation  for  Research,  Technology  and  Development  is  gratefully  acknowledged.
\end{acks}

\bibliographystyle{ACM-Reference-Format}
\bibliography{napire}

\appendix

\medskip
\section{List of Contributors and Roles}

Table~\ref{tab:authorshipdetails} introduces the details of the authorships with respect to the roles taken in the NaPiRE project and in context of this particular manuscript. To this end, we introduce a role concept based on the classification scheme as discussed by Brand \textit{et al}.~\cite{Brand15}. In our context, we distinguish the following roles\footnote{Those roles marked with an $*$ are the exact same as introduced in~\cite{Brand15}.}:
\begin{compactitem}
\item Conceptualisation$^{*}$: Ideas; formulation or evolution of overarching research goals and aims.
\item Project Administration$^{*}$: Management and coordination responsibility for the research activity planning and execution.
\item Methodology$^{*}$: Development or design of methodology; creation of models.
\item Instrument Design: Development / re-design of the instrument used in this replication. 
\item Data Collection: Data collection as national representative in the respective country using the provided infrastructure.
\item Data Analysis: Application of analysis techniques to study and interpret the data.
\item Data Curation$^{*}$: Management activities to annotate (produce metadata), scrub data and maintain research data (including software code, where it is necessary for interpreting the data itself) for initial use and later reuse.
\item Data Visualisation$^{*}$: Preparation, creation and/or presentation of the published work, specifically visualisation/ data presentation.
\item Writing - Original Draft$^{*}$: Preparation, creation and/or presentation of the published work, specifically writing the initial draft (including substantive translation).
\item Data - Review \& Editing$^{*}$: Preparation, creation and/or presentation of the published work by those from the original research group, specifically critical review, commentary or revision -- including pre- or post-publication stages
\end{compactitem}

The first two authors are the initiators and coordinators of the overall initiative. Further, we formed a group of lead authors for this particular manuscript (the first seven authors) to do the data analysis and / or the writing process.  

\begin{table}[htb]
\scriptsize
\centering
\caption{Authorship details}
\label{tab:authorshipdetails}
\begin{tabular}{p{0.45\linewidth}cccccccccc}
\toprule
\textbf{Author}  & \rot{\textbf{Conceptualisation}} & \rot{\textbf{Project Administration}} & \rot{\textbf{Methodology}} & \rot{\textbf{Instrument Design}} & \rot{\textbf{Data Collection}} & \rot{\textbf{Data Analysis}} & \rot{\textbf{Data Curation}} & \rot{\textbf{Data Visualisation}} & \rot{\textbf{Writing - Draft}} & \rot{\textbf{Writing - Review \& Editing}}\\ \midrule
S. Wagner & X & X & X & X & X & X & & X & X & X\\
D. M\'{e}ndez Fern\'{a}ndez & X & X & X & X &X & X & X & X& X & X \\
M. Felderer & & & & X& X&X &  & X &X & X\\
A. Vetr\`{o} & & & X && & X &X & X& & X\\
M. Kalinowski & & & &X&X & X& X & X &X & X\\
R. Wieringa & & & X &X&X & & & & &X \\
D. Pfahl & & & & X& X& & & &  &X\\
T. Conte & & & &&X & & & & &X\\
M.-T. Christiansson  & && & &X & & & & &X\\
D. Greer & & & &X& X& & & &&X\\
C. Lassenius & & & & X& X& & & & &X\\
T. M\"annist\"o & & & & X&X & &&  & &X\\
M. Nayebi & & & & & X& & & & &X\\
M. Oivo & & & &X& X& & & &&X\\
B. Penzenstadler & & & & &X & & & & &X\\
R. Prikladnicki & & & & &X &  & & & &X \\
G. Ruhe & & &  && X& & & & &X\\
A. Schekelmann & & & & &X & & & & &X  \\
S. Sen & & & & & X& & & & &X\\
R. Spinola & & & & &X &  &  & & &X\\
A. Tuzcu & & & & & & & & & &X \\
J. L. de la Vara & & & & &X & & & & &X\\
D. Winkler & & & & & & & & & X &X\\
\bottomrule
\end{tabular}
\end{table}

\clearpage
\section{Results of K-W tests}
\label{sec:appendix-k-w-tests}

\begin{table}[htb]
\footnotesize
\centering
\caption{Kruskall-Wallis test for evaluating country effect on results: part 1\label{tab:kwtestpart1}}{
\begin{tabular}{lp{3cm}lr}
\toprule
\textbf{No.} & \textbf{Question} & \textbf{Option} & \textbf{p-value} \\ \midrule
Q1 & \multicolumn{2}{l}{What is the size of your company?}  & \textbf{0.035} \\
Q3 & \multicolumn{2}{l}{Does your company participate in globally distributed projects?} & \textbf{\textless0.001} \\
Q5 & \multicolumn{2}{l}{To which project role are you most frequently assigned?} & 0.053 \\
Q6 & \multicolumn{2}{l}{How do you rate your experience in this role?} & 0.173 \\
Q7 & \multirow{4}{3cm}{Which org.\ role does your company take most frequently in your projects?} & Customer & 0.116 \\
 &  & Product development & \textbf{0.010} \\
 &  & Contractor & \textbf{0.003} \\
 &  & Other & 0.743 \\
Q8 & \multirow{3}{3cm}{Which process model do you follow (or a variation of it)?} & Waterfall & 0.690 \\
 &  & V-Modell XT & 0.105 \\
 &  & Scrum & 0.418 \\
 &  & Extreme Programming (XP) & 0.276 \\
 &  & Rational Unified Process & \textbf{0.034} \\
 &  & Other & 0.195 \\
Q9 & \multirow{2}{3cm}{How do you elicit requirements?} & Interviews & \textbf{0.010} \\
 &  & Scenarios & \textbf{0.038} \\
 &  & Prototyping & 0.555 \\
 &  & Facilitated meetings (including workshops) & \textbf{0.012} \\
 &  & Observation & \textbf{0.002} \\
Q10 & \multirow{3}{3cm}{How do you document functional requirements?} & Free form textual (Domain/Business Process Models) & 0.743 \\
 &  & Free form textual (Use Case Models) & 0.135 \\
 &  & Free form textual (Goal Models) & 0.561 \\
 &  & Free form textual (Data Models) & 0.629 \\
 &  & Free form textual (Structured Requirements Lists)& 0.654 \\
 &  & Textual with constraints (Domain/Business Process Models) & 0.500 \\
 &  & Textual with constraints (Use Case Models) & 0.975 \\
 &  & Textual with constraints (Goal Models) & 0.176 \\
 &  & Textual with constraints (Data Models) & 0.226 \\
 &  & Textual with constraints (Structured Requirements Lists) & 0.495 \\
 &  & Semi-formal (UML) (Domain/Business Process Models) & \textbf{0.046} \\
 &  & Semi-formal (UML) (Use Case Models) & \textbf{0.041} \\
 &  & Semi-formal (UML) (Goal Models) & 0.972 \\
 &  & Semi-formal (UML) (Data Models) & 0.100 \\
 &  & Semi-formal (UML) (Structured Requirements Lists) & 0.853 \\
 &  & Formal  (Domain/Business Process Models) & 0.572 \\
 &  & Formal  (Use Case Models) & \textbf{0.015} \\
 &  & Formal  (Goal Models) & 0.068 \\
 &  & Formal  (Data Models) & 0.914 \\
 &  & Formal  (Structured Requirements Lists) & 0.518 \\
Q11 & \multicolumn{2}{l}{How do you document non-functional requirements?} & 0.304 \\
Q12 & \multicolumn{2}{l}{How do you deal with changing requirements after the initial release?} & 0.550 \\
Q13 & \multirow{2}{3cm}{Which traces do you explicitly manage?} & Traces between requirements and code. & \textbf{0.001} \\
 &  & Traces between requirements and design documents. & 0.558 \\
 &  & None. & 0.180 \\
Q14 & \multirow{3}{3cm}{How do you analyse the effect of changes to requirements?} & We do impact analysis between requirements. & \textbf{0.023} \\
 &  & We do impact analysis on the code. & 0.255 \\
 &  & We do not analyse the effect of changes to requirements. & 0.070 \\
Q15 & \multirow{3}{3cm}{How do you align the software test with the requirements?} & Testers participate in requirements reviews. & 0.130 \\
 &  & We check the coverage of requirements with tests. & 0.061 \\
 &  & We define acceptance criteria for requirements. & \textbf{0.022} \\
 &  & We derive tests from system models. & 0.557 \\
 &  & We do not allign test and requirements. & 0.532 \\ \bottomrule
\end{tabular}}
\end{table}

\begin{table}[htb]
\footnotesize
\centering
\caption{Kruskall-Wallis test for evaluating country effect on results: part 2\label{tab:kwtestpart2}}{
\begin{tabular}{lp{3cm}p{8cm}l}
\toprule
\textbf{No.} & \textbf{Question} & \textbf{Option} & \textbf{p-value} \\
\midrule
Q16 & \multirow{3}{3cm}{What RE standard have you established at your company?} & A standard that is predefined according to a regulation (e.g.. ITIL) & \textbf{\textless0.001} \\
 &  & A standard that is predefined by the development process (e.g.. Rational Unified Process. Scrum) & 0.215 \\
 &  & An own standard that defines the coarse process with deliverables. milestones. and phases & 0.192 \\
 &  & An own standard that defines the process including roles and responsibilities. & 0.610 \\
 &  & An own standard that defines artefacts and offers document templates & 0.095 \\
 &  & None & \textbf{0.029} \\
Q17 & \multirow{5}{3cm}{Which reasons do you agree with as a motivation to define a company standard for RE in your company?} & Compliance to regulations and standards (like CMMI) & 0.140 \\
 &  & Seamless development by integrating Requirements Engineering into the development process & 0.344 \\
 &  & Better tool support & 0.361 \\
 &  & Formal prerequisite for project acquisition in your domain & 0.572 \\
 &  & Support of distributed development & 0.284 \\
 &  & Better support of progress control & 0.680 \\
 &  & Better quality assurance of the artefacts (e.g.. within quality gates) & 0.169 \\
 &  & Support of benchmarks and / or comparison of different projects & 0.466 \\
 &  & Support of project management and planning & 0.753 \\
 &  & Higher efficiency & 0.456 \\
 &  & Knowledge transfer & 0.827 \\
Q18 & \multirow{5}{3cm}{Which reasons do you see as a barrier to define a company standard for RE in your company?} & Higher process complexity & 0.694 \\
 &  & Higher demand for communication & 0.549 \\
 &  & Lower efficiency & \textbf{0.002} \\
 &  & Missing willingness for changes & 0.074 \\
 &  & Missing possibilities of standardisation & 0.066 \\
Q19 & \multicolumn{2}{l}{Is the requirements engineering standard mandatory and practiced?} & \textbf{0.015} \\
Q20 & \multirow{4}{3cm}{How do you check the application of your requirements engineering standard?} & Via project assessments & 0.092 \\
 &  & Via analytical quality assurance. e.g.. as part of quality gates & 0.366 \\
 &  & Via constructive quality assurance. e.g.. via checklists or templates & 0.252 \\
 &  & It is not checked. & 0.204 \\
Q21 & \multirow{4}{3cm}{How do you perform change management in your requirements engineering?} & We have a continuous change management. & \textbf{0.010} \\
 &  & We have a change management approach that applies after formally accepting a requirements specification. & 0.701 \\
 &  & We have a change management that applies during RE. & 0.070 \\
 &  & We do not consider a change management in RE. & \textbf{0.009} \\
Q22 & \multirow{4}{3cm}{How is your RE standard applied (tailored) in your regular projects?} & We have defined a tailoring approach that continuously guides the application of the standard in our project & 0.161 \\
 &  & We have tool support for tailoring our Requirements Engineering standard & 0.164 \\
 &  & At the beginning of a project. the project lead / requirements engineer tailors the standard based on experiences & \textbf{0.018} \\
 &  & We do not consider a particular tailoring approach & \textbf{0.025} \\
Q23 & \multicolumn{2}{l}{Is your RE continuously improved?} & \textbf{0.035} \\
Q24 & \multirow{4}{3cm}{Why do you continuously improve your requirements engineering?} & It helps us to determine our strenghts and weaknesses and act accordingly & 0.604 \\
 &  & An improvement is expected by our customer & 0.330 \\
 &  & We conduct the improvement to obtain a certain certification. & \textbf{0.004} \\
 &  & An improvement is demanded by a regulation (e.g.. CMMI. Cobit. or ITIL) & \textbf{0.010} \\
Q25 & \multicolumn{2}{l}{Do you use a normative, external standard for your improvement?} & \textbf{0.002}\\ \bottomrule
\end{tabular}}
\end{table}
\clearpage

\end{document}